\documentclass[11pt,a4paper]{article}
\pdfoutput=1
\usepackage{graphicx}
\usepackage{amssymb,amsmath,mathtools,mathrsfs} % maths
\usepackage[normalem]{ulem} % editing
\usepackage{enumerate,todonotes} 
\usepackage{jcappub}
\usepackage{placeins}
\usepackage{multirow}
\usepackage{hyperref}
\usepackage{float}
\usepackage{setspace}
\usepackage{graphics}
\usepackage{fullpage}
\usepackage{epstopdf}
\usepackage{comment}
\usepackage[utf8]{inputenc}

\makeindex
\begin{document}
\hypersetup{pageanchor=false} 
\title{New massive bigravity cosmologies with double matter coupling}

\author[a,b]{Macarena Lagos,}
\author[a]{Johannes Noller}
\affiliation[a]{Astrophysics, University of Oxford, DWB,\\
Keble road, Oxford OX1 3RH, UK}
\affiliation[b]{Theoretical physics, Blackett Laboratory, Imperial College London,\\
Prince Consort Road, London SW7 2BZ, UK}

\emailAdd{m.lagos13@imperial.ac.uk}
\emailAdd{noller@physics.ox.ac.uk}

\abstract{We study a previously largely unexplored branch of homogeneous and isotropic background solutions in ghost-free massive bigravity with consistent double matter coupling. For a certain family of parameters we find `self-inflated' FLRW cosmologies, i.e.~solutions with an accelerated early-time period during the radiation-dominated era. In addition, these solutions also display an accelerated late-time period closely mimicking GR with a cosmological constant. Interestingly, within this family, the particular case of $\beta_1=\beta_3=0$ gives bouncing cosmologies, where there is an infinite contracting past, a non-zero minimum value of the scale factor at the bounce, and an infinite expanding future.}


\maketitle

%--------------------------------------------------------------------------------------------------------------------------------------------

\section{Introduction}

Our present cosmological standard model -- $\Lambda$CDM and with it General Relativity (GR) -- still faces several problems. These range from the technical un-naturalness of the cosmological constant $\Lambda$ to requiring the presence of further unknown `dark' components, namely dark energy (if not fully accounted for by $\Lambda$) and dark matter, making up nearly $95\%$ of the total effective matter content of the universe \cite{Ade:2015xua}, if GR is indeed the correct theory of gravity on all scales.
Massive bigravity, as proposed in \cite{Hassan:2011hr, Hassan:2011zd}, is a natural extension to massive gravity \cite{deRham:2010ik, deRham:2010kj, Hassan:2011vm}, and a promising alternative to GR, which may help to answer some of the problems mentioned above. This model is an interacting bimetric theory containing more physical degrees of freedom (DoF) than GR, as it propagates two spin-2 fields, corresponding to one massive (5 DoF) and one massless (2 DoF) graviton, in contrast to the single massless graviton of GR. 

Massive bigravity passes several of the immediate consistency checks any theory of gravity needs to pass. It agrees with solar system constraints on the presence of a fifth force \cite{Volkov:2012wp,Enander:2013kza,Babichev:2013pfa} (the massive graviton would propagate such an extra force) via the Vainshtein mechanism \cite{Vainshtein:1972sx}. On cosmological scales, it also gives rise to viable homogeneous and isotropic (FLRW) background solutions \cite{Akrami:2012vf,Akrami:2013pna,Konnig:2013gxa,Enander:2014xga}. The specific massive bigravity interactions (between both metrics) were carefully chosen to avoid an extra propagating scalar DoF, namely a Boulware-Deser ghost \cite{Boulware:1973my,deRham:2010ik,deRham:2010kj,Hassan:2011vm,Hassan:2011hr,deRham:2011rn, Hassan:2011ea,Hassan:2012qv}, causing the theory to be unstable. However, the absence of the Boulware-Deser ghost does not guarantee the model to be instability free, as some of the DoF of the gravitons may still behave as ghosts, tachyons or create gradient instabilities. In fact, this is generically the case, as has been shown at the level of perturbations in previous cosmological studies (see below). 

The cosmology of ghost-free massive bigravity has been extensively studied when matter is coupled to one of the metrics only (single matter coupling) \cite{Volkov:2011an,vonStrauss:2011mq,Comelli:2011zm,Comelli:2012db,Berg:2012kn,Akrami:2012vf,Sakakihara:2012iq,Akrami:2013pna,DeFelice:2013nba,Fasiello:2013woa,Volkov:2013roa,Konnig:2013gxa,Berezhiani:2013dw,Konnig:2014dna,Solomon:2014dua,Comelli:2014bqa,Konnig:2014xva,Lagos:2014lca,Cusin:2014psa,Enander:2015vja,Nersisyan:2015oha,Soloviev:2015wya,Amendola:2015tua,Johnson:2015tfa,Konnig:2015lfa,Cusin:2015pya,Fasiello:2015csa}. In this setting, two main branches of solutions were identified. Even though at the level of the background both branches could lead to viable cosmologies, at the level of linear perturbations the so-called expanding branch (also known as finite branch) was found to have gradient instabilities on the scalar sector, leading to an exponential growth of these perturbations \cite{Comelli:2012db}, and in turn breaking the validity of perturbation theory at early times\footnote{Some papers have suggested ways to overcome this issue \cite{Akrami:2015qga,Mortsell:2015exa}, though.}. On the other hand, the so-called bouncing branch (also known as infinite branch) was found to have the helicity-0 mode of the massive graviton behaving as a ghost, and has tachyon instabilities in the tensor sector \cite{Lagos:2014lca,Cusin:2014psa,Amendola:2015tua,Johnson:2015tfa,Konnig:2015lfa,Cusin:2015pya}\footnote{General relativity with a cosmological constant is of course a particular limit of ghost-free bigravity and massive gravity theories and as such one would naively expect that a GR-like and hence instability-free evolution can be recovered by suppressing any non-GR operators, i.e.~by establishing a hierarchy between coupling constants in the theory. Recent studies have confirmed that this is indeed the case. For instance, in a study of the expanding branch in \cite{Akrami:2015qga}, it was found that for appropriate choices of parameters, the scalar exponential instability could be moved outside the regime of validity of the theory (in an effective field theory sense), or equivalently to `very early times'. When curing the instabilities of the theory in this way, the price to pay is of course suppressing any non-GR signatures.}.

More general consistent theories of metrics/spin-2 fields beyond massive (bi-)gravity with a single matter coupling have also been explored recently. New kinetic interactions were investigated in \cite{Folkerts:2011ev,Hinterbichler:2013eza,Kimura:2013ika,deRham:2013tfa,
Gao:2014jja,Noller:2014ioa,deRham:2015rxa}, generalisations of the potential interactions of massive bigravity to $N$ multiple metrics in \cite{Khosravi:2011zi,Hinterbichler:2012cn,Hassan:2012wt,Nomura:2012xr,Tamanini:2013xia,Noller:2013yja,Scargill:2014wya,Goon:2014paa,Noller:2015eda}, and new couplings to matter in \cite{Baccetti:2012re,Hassan:2012wr,Capozziello:2012re,Akrami:2013ffa,Tamanini:2013xia,Bamba:2013hza,Aoki:2013joa,Comelli:2014bqa,Akrami:2014lja,DeFelice:2014nja,Yamashita:2014fga,deRham:2014naa,Noller:2014sta,Enander:2014xga,Hassan:2014gta,Gumrukcuoglu:2014xba,deRham:2014fha,Schmidt-May:2014xla,Aoki:2014cla,Gao:2014xaa,Hinterbichler:2015yaa,Blanchet:2015sra,Aoki:2015xqa,Bernard:2015gwa}. These matter couplings allow matter to couple to both metrics and we will therefore refer to them as `double matter couplings'. Generic such couplings re-introduce the Boulware-Deser ghost at an unacceptably low scale, however the specific couplings of \cite{deRham:2014naa,Noller:2014sta} stand out in that they are consistent ghost-free double matter couplings. 
Note that this statement is taken in an effective field theory sense. They are the only known non-derivative couplings that do not introduce a ghost up to the $\Lambda_3$ strong coupling scale (for a discussion of derivative couplings see \cite{Heisenberg:2015wja}), so massive bigravity with such a double matter coupling can be safely considered as an effective field theory with a cutoff at $\Lambda_3$ or above. A ghost is `present' at larger energies, which are beyond the regime of validity of the theory (in fact the ghost scale may set the theory's cutoff), but the ghost is of course not at all present in the low-energy effective theory below the cutoff. In fact it has recently been argued that the couplings of \cite{deRham:2014naa,Noller:2014sta} are the unique such couplings for a cutoff $\ge \Lambda_3$ \cite{deRham:2015cha,Huang:2015yga,Heisenberg:2015iqa}.

In the context of this double coupling, some homogeneous and isotropic cosmological solutions were studied in \cite{Enander:2014xga}, where viable (background) evolutions were found. However, at the level of linear perturbations, tachyonic, gradient, and ghost instabilities were found for these solutions \cite{Comelli:2015pua, Gumrukcuoglu:2015nua} in the tensor, vector and scalar sector, respectively. 
In the light of these results, in this paper we analyse previously unexplored homogeneous and isotropic background solutions with the consistent double coupling mentioned above. Specifically, we find new self-inflated cosmologies, by which we mean solutions with an accelerated early-time period during the radiation-dominated era. Depending on the choice of parameters, this inflating period could happen within the regime of validity of massive bigravity, and thus give an interesting alternative to standard cosmic inflation, where, contrary to GR, no unknown inflationary field would need to be introduced. However, even in the case where the inflationary early universe phase is outside of the regime of validity of the theory, the new solutions we find also give new consistent background cosmologies for massive bigravity at late times, closely mimicking GR with a cosmological constant. In addition, all the self-inflated solutions found have a minimum non-zero value of the scale factor (and associated finite energy densities) and thus avoid any physical Big Bang singularity. The particular case $\beta_1=\beta_3=0$ gives a bouncing universe, i.e.~where there is an infinite contracting past, followed by a bounce with an infinite expanding future. A detailed analysis on the stability of linear perturbations around these backgrounds is beyond the scope of this paper, but we argue that some improvements are expected compared to previous solutions found in \cite{Comelli:2015pua, Gumrukcuoglu:2015nua}. Specifically, we find that for appropriate values of the parameters of the model, gradient instabilities of vector perturbations can be avoided.
\\

\noindent {\it Outline}: The outline of this paper is as follows. In Section \ref{Sec1:BigravityDoublyCoupled}, we introduce massive bigravity with a consistent double coupling to matter. In section \ref{Sec2:Background}, we assume a flat, homogeneous and isotropic universe, and describe some properties of the solutions in two possible different branches, which come from two solutions to the Bianchi constraint of the theory. In Section \ref{Sec3:BranchI}, we focus on the previously less explored of the two branches. We show specific analytical approximate solutions for this branch (as introduced in Section \ref{Sec2:Background}) in some relevant regimes (early times, bouncing period, and late times) for different sets of parameters, as well as numerical solutions. Finally, in Section \ref{Sec:Conclusions} we summarise our results, discuss consequences and future directions. 
\\

\noindent {\it Conventions}: Throughout this paper Greek indices such as $\mu, \nu$ denote spacetime indices. Indices of tensors depending on one metric only will be raised and lowered with their corresponding metric, whereas raising and lowering procedures for quantities depending on more than one metric will be explicitly specified where required. The Einstein summation convention is implied as usual. In addition, bracketed indices $(g)$, and $(f)$, label tensors to clarify their relation to/dependence on the two different metrics $g_{\mu\nu}$ and $f_{\mu\nu}$ to be considered. These label indices are not automatically summed over and whether they are upper or lower indices carries no meaning. Finally, we will be using Planck units.

%---------------------------------------------------------------------------------------------------------------------------------------
\section{Bigravity model with double matter coupling}\label{Sec1:BigravityDoublyCoupled}

Let us start with the massive bigravity action as proposed in \cite{Hassan:2011zd}:
\begin{equation}\label{MGaction}
S= \frac{M_g^2}{2}\int d^4x \sqrt{-g}R_{(g)} +\frac{M_f^2}{2}\int d^4x \sqrt{-f}R_{(f)} - m^4\int d^4x \sqrt{-g}\sum_{n=0}^4\beta_n e_n\left(\sqrt{g^{-1}f}\right)+S_\text{m},
\end{equation}
which includes two Einstein-Hilbert terms for two metrics $g_{\mu\nu}$ and $f_{\mu\nu}$, with mass scales given by $M_g$ and $M_f$, and their associated Ricci scalars $R_{(g)}$ and $R_{(f)}$, respectively. Throughout this paper we will use $M_g=M_f=M_{\rm pl}$, where $M_{\rm pl}$ corresponds to the Planck mass. In addition, these two metrics have a very particular interaction term which includes a mass scale $m$, five dimensionless parameters $\beta_n$, and the functions $e_n(\mathbb{X})$, which correspond to the elementary symmetric polynomials of the eigenvalues of the matrix $\mathbb{X}\equiv\sqrt{g^{-1}f}$, which in turns satisfies $(\mathbb{X}^2)^{\mu}{}_{\nu}=g^{\mu\alpha}f_{\alpha\nu}$. As previously mentioned, this interaction term is the only way to non-dynamically couple these two metrics without introducing a Boulware-Deser ghost. Finally, we also include a matter coupling $S_\text{m}$ of the form 
\begin{equation}\label{EqSm}
S_{\rm m} = \int d^4x \sqrt{-g^{\rm eff}}{\cal L}_{\rm m}[\Phi_i,g_{\mu \nu}^{\rm eff}],
\end{equation}
where ${\cal L}_{\rm m}$ is the matter Lagrangian and all matter fields $\Phi_i$ are minimally coupled to a single effective metric $g_{\mu\nu}^{\rm eff}$ in accordance with the weak equivalence principle. Metric $g_{\mu\nu}^{\rm eff}$ is in general a function of both metrics in the theory, $g_{\mu\nu}$ and $f_{\mu\nu}$, and it will correspond to the physical metric describing the space-time, as from eq.~(\ref{EqSm}) we notice that matter will follow geodesics of this effective metric.

In this paper we focus on the double matter coupling proposed by \cite{deRham:2014naa,Noller:2014sta}, working in the metric formulation used by \cite{deRham:2014naa} enabling easier comparison with standard cosmology. In this double coupling, the effective metric $g^{\text{eff}}_{\mu\nu}$ is given by\footnote{Note that in the vielbein formulation the effective metric vielbein takes a remarkably simple form: a linear superposition of the vielbeins for $g_{\mu\nu}$ and $f_{\mu\nu}$ \cite{Noller:2014sta}.}:
\begin{equation}\label{gEff}
g^{\text{eff}}_{\mu\nu}=\alpha^2 g_{\mu\nu}+ 2 \alpha\beta g_{\mu\alpha} \mathbb{X}^\alpha{}_\nu +\beta^2f_{\mu\nu},
\end{equation}
where $\alpha$ and $\beta$ are two dimensionless arbitrary parameters. Notice that singly-coupled cases are recovered by setting either $\alpha$ or $\beta$ to zero. As previously mentioned, massive bigravity with matter coupled to the effective metric given by eq.~(\ref{gEff}) does not introduce the Boulware-Deser ghost at least up to the strong coupling scale $\Lambda_3=(m^3M_{\rm pl})^{1/3}$, and thus this theory can safely be considered as an effective field theory with a cutoff at $\Lambda_3$ or above. In addition, this double coupling was studied in the context of massive gravity (only one dynamical metric), where flat FLRW solutions were found \cite{deRham:2014naa}. These solutions identified qualitatively new massive gravity cosmologies, as no exact flat FLRW solutions are present with single couplings \cite{D'Amico:2011jj} (although solutions with cosmologies arbitrarily close to FLRW are possible). Furthermore, it was found that the Boulware-Deser ghost does not propagate at all at the level of linear perturbations around FLRW backgrounds \cite{deRham:2014naa}.\footnote{The mentioned double coupling has other interesting features as well. For instance, one-loop corrections from matter loops do not detune the particular ghost-free structure of the interaction term of the theory, and thus the Boulware-Deser ghost is not re-introduced by these corrections \cite{deRham:2014naa,Heisenberg:2014rka}. Note that this result is expected to hold for arbitrary matter loop order, as matter loop corrections should sum up to an effective cosmological constant as discussed in \cite{Noller:2014sta}.} 

In the bimetric context, action (\ref{MGaction}) and the double matter coupling with the effective metric given by eq.~\eqref{gEff} have the property of being invariant under the following transformation: 
\begin{equation}\label{Transf}
g_{\mu\nu} \leftrightarrow f_{\mu\nu}, \quad \beta_n \leftrightarrow \beta_{4-n},\quad M_g \leftrightarrow M_f,\quad \alpha \leftrightarrow \beta,
\end{equation}
and therefore, contrary to the singly-coupled models (where any potential symmetry between the metrics is broken by the matter coupling), both metrics here have a similar role. 

The equations of motion of the bimetric action (\ref{MGaction}) with the previously mentioned double matter coupling are \cite{Schmidt-May:2014xla}: 
\begin{align}
& \sqrt{-g}\tilde{G}_{(g)\rho\sigma}(F^{\rho\mu}g^{\sigma\nu}+F^{\rho\nu}g^{\sigma\mu})= \sqrt{-g^\text{eff}}T^{\rho\sigma}\alpha(2\beta\delta^{\mu}_\rho\delta^\nu_\sigma+\alpha f^{\mu\lambda}F_{\lambda\sigma}\delta^\nu_\rho+\alpha f^{\nu\lambda}F_{\lambda\sigma}\delta^\mu_\rho),\label{Eqg}\\
& \sqrt{-f}\tilde{G}_{(f)\rho\sigma}(F^{\rho\mu}f^{\sigma\nu}+F^{\rho\nu}f^{\sigma\mu})= \sqrt{-g^\text{eff}}T^{\rho\sigma}\beta(2\alpha\delta^{\mu}_\rho\delta^\nu_\sigma+\beta g^{\mu\lambda}F_{\lambda\sigma}\delta^\nu_\rho+\beta g^{\nu\lambda}F_{\lambda\sigma}\delta^\mu_\rho)\label{Eqf},
\end{align}
where $\tilde{G}_{(g)\mu\nu}$ and $\tilde{G}_{(f)\mu\nu}$ are the modified gravitational equations in the absence of matter for the metrics $g_{\mu\nu}$ and $f_{\mu\nu}$, respectively. Explicitly, $\tilde{G}_{(g)\mu\nu}$ and $\tilde{G}_{(f)\mu\nu}$ are given by:
\begin{align}
& \tilde{G}_{(g)\mu\nu} = R_{(g)\mu\nu}-\frac{1}{2}g_{\mu\nu}R_{(g)} +\frac{m^4}{2}\sum_{n=0}^3(-1)^n\beta_n\left[g_{\mu\lambda}Y^\lambda_{(n)\nu}\left(\sqrt{g^{-1}f}\right)+ g_{\nu\lambda}Y^\lambda_{(n)\mu}\left(\sqrt{g^{-1}f}\right)\right],\label{Gg}\\
 & \tilde{G}_{(f)\mu\nu} = R_{(f)\mu\nu}-\frac{1}{2}f_{\mu\nu}R_{(f)} + \frac{m^4}{2}\sum_{n=0}^3(-1)^n\beta_{4-n}\left[f_{\mu\lambda}Y^\lambda_{(n)\nu}\left(\sqrt{f^{-1}g}\right)+f_{\nu\lambda}Y^\lambda_{(n)\mu}\left(\sqrt{f^{-1}g}\right)\right],\label{Gf}
\end{align}
where the functions $Y^\lambda_{(n)\nu}(\mathbb{X})$, written in matrix notation, are given by:
\begin{align}
Y_{(0)}=&\mathbb{I},\nonumber\\
Y_{(1)}=&\mathbb{X}-\mathbb{I}[\mathbb{X}],\nonumber\\
Y_{(2)}=&\mathbb{X}^2-\mathbb{X}[\mathbb{X}]+\frac{1}{2}\mathbb{I}\left([\mathbb{X}]^2-[\mathbb{X}^2]\right),\nonumber\\
Y_{(3)}=&\mathbb{X}^3-\mathbb{X}^2[\mathbb{X}]+\frac{1}{2}\mathbb{X}\left([\mathbb{X}]^2-[\mathbb{X}^2]\right) -\frac{1}{6}\mathbb{I}\left([\mathbb{X}]^3-3[\mathbb{X}][\mathbb{X}^2]+2[\mathbb{X}^3]\right),
\end{align}
where $\mathbb{I}$ is the identity matrix and $[\mathbb{X}]$ stands for the trace of the matrix $\mathbb{X}$. Here, it is understood that $(n)$ is a label taking the values $n=0,1,2,3,4$. In addition, in eq.~(\ref{Eqg})-(\ref{Eqf}) we have defined $F_{\mu\nu}\equiv g_{\mu\alpha}\mathbb{X}^{\alpha}{}_{\nu}$ with inverse $F^{\mu\nu}$ such that $F^{\mu\alpha}F_{\alpha\nu}=\delta^{\mu}_\nu$, and the stress-energy tensor $T^{\mu\nu}$ as:
\begin{equation}
T^{\mu\nu}\equiv \frac{2}{\sqrt{-g^\text{eff}}}\left(\frac{\delta S_m}{\delta g^\text{eff}_{\mu\nu}}\right),
\end{equation}
where it is understood that the indices of $T^{\mu\nu}$ are lowered and raised using the effective metric $g^\text{eff}_{\mu\nu}$. In this sense our $T^{\mu\nu}$ is really a $T^{\mu\nu}_{\rm eff}$, but we drop the ${\rm eff}$ label in order to avoid clutter. 

We remark that massive bigravity with double matter coupling is invariant under (a single copy of) general coordinate transformations but, contrary to GR, deriving the covariant conservation of the stress-energy tensor with respect to the effective matter metric from the Bianchi constraints is not as straightforward. In this paper we will be working with matter (perfect fluids, scalar fields) for which this conservation is automatically satisfied, i.e.~where
\begin{equation}\label{EqDT}
\nabla_{\mu}^\text{eff}T^{\mu\nu}=0
\end{equation}
is a direct consequence of the matter equations of motion and where $\nabla_{\mu}^\text{eff}$ corresponds to the covariant derivative with respect to $g^\text{eff}_{\mu\nu}$. For more general types of matter an extra assumption may need to be imposed for \eqref{EqDT} to hold true. The Bianchi constraints will nevertheless play a special role, as we will see in the next section. Specifically, in massive bigravity with double matter coupling, the Bianchi constraints lead to a consistency equation that identifies two qualitatively different branches of solutions.

%-----------------------------------------------------------------------------------------------------------------------------------------------

\section{Cosmological background}\label{Sec2:Background}
In this section we focus on cosmological predictions of the model presented in the previous section. In particular, we show the main equations describing the evolution of a flat homogeneous and isotropic universe with a perfect fluid. We also show two possible branches of solutions allowed by the Bianchi constraint, as well as some of their properties.

We assume flat FLRW metrics. Explicitly, both metrics $f_{\mu\nu}$ and $g_{\mu\nu}$ will be written in conformal time $\tau$ as:
\begin{align}
 ds_g^2 &= a_g^2\left[-X_g^2 d\tau^2+\delta_{ij}dx^i dx^j\right], \label{sg}\\
 ds_f^2 &= a_f^2\left[-X_f^2d\tau^2+\delta_{ij}dx^i dx^j\right]. \label{sf}
\end{align}
By means of eq.~(\ref{gEff}), we find that the physical space-time metric $g^\text{eff}_{\mu\nu}$ shares the same symmetries and therefore can be written in the same form:
\begin{equation}
ds^2_{\text{eff}}= a^2\left(-X^2d\tau^2+\delta_{ij}dx^i dx^j\right), \label{sgeff}
\end{equation}
where the scale factor $a$ and the shift function $X$ are related to those of the metrics $g_{\mu\nu}$ and $f_{\mu\nu}$ by\footnote{The fact that scale factors linearly superpose in this way is a direct consequence of the effective metric vielbein being a linear superposition of the vielbeins in the theory.}:
\begin{align}
a &= \alpha a_g+\beta a_f, \label{aEffective}\\
X &= \frac{\alpha a_gX_g+\beta a_fX_f}{a}. \label{XEffective}
\end{align}
Note that the conformal time $\tau$ will be related to the physical time $t$ by $aX d\tau= dt$. From eq.~(\ref{aEffective}) we find that the comoving Hubble ratio $\mathcal{H}$ of the space-time metric $g^{\rm eff}_{\mu\nu}$ is related to those of the metrics $g_{\mu\nu}$ and $f_{\mu\nu}$ by:
\begin{equation}\label{Friedmann}
\mathcal{H}= \frac{\alpha \mathcal{H}_g + \beta N\mathcal{H}_f}{(\beta N+\alpha)},
\end{equation}
where we have defined $N \equiv a_f/a_g$, $\mathcal{H} \equiv a'/a$, and $\mathcal{H}_i \equiv a_i'/a_i$ (with $i=(g,f)$).
In addition, we couple $g^\text{eff}_{\mu\nu}$ to a perfect fluid with a stress-energy tensor given by:
\begin{equation}\label{MatterAnsatz}
T^{\mu}{}_{\nu}=(p+\rho)u^{\mu}u_{\nu}+p\delta^{\mu}_{\nu},
\end{equation}
where $p$ is the pressure of the fluid, $\rho$ its energy density, and $u^{\mu}=((aX)^{-1},0,0,0)$ its isotropic 4-velocity vector. 

By substituting eq.~(\ref{sg})-(\ref{sf}) and eq.~(\ref{MatterAnsatz}) into eq.~(\ref{Eqg})-(\ref{Eqf}) we get the following equations of motion determining the evolution a flat homogeneous and isotropic universe:
\begin{align}
\mathcal{H}_g^2&=\frac{X_g^2a_g^2}{3}\left[\alpha \rho\frac{a^3}{a_g^3}+ m^4 (\beta_0+3N\beta_1+3N^2\beta_2+N^3\beta_3) \right],\label{Friedg}\\
\mathcal{H}_g' &=\frac{1}{2}\left[2 \mathcal{H}_g \frac{ X_g'}{ X_g }+ 2\mathcal{H}_g^2+ m^4 NZa_g^2 X_g (X_f-X_g) + \frac{ a^4X_g(\rho X_g+pX)(X- X_f)  }{(X_f-X_g) a_g^2} \right],\label{DHg}\\
\mathcal{H}_f^2&=\frac{X_f^2a_f^2}{3}\left[ \beta \rho\frac{a^3}{a_f^3}+m^4 (\beta_1N^{-3}+3N^{-2}\beta_2+3N^{-1}\beta_3+\beta_4)\right],\label{Friedf}\\
\mathcal{H}_f' &= \frac{1}{2}\left[2 \mathcal{H}_f \frac{ X_f'}{ X_f }+ 2\mathcal{H}_f^2+ \frac{ m^4 Za_f^2 X_f (X_g-X_f) }{N^{3} }+ \frac{ a^4X_f(\rho X_f+pX)(X- X_g)  }{(X_g-X_f) a_f^2 } \right],\label{DHf}
\end{align}
where we have chosen $\mathbb{X}=\mbox{diag}(X_fN/X_g,N,N,N)$\footnote{The massive bigravity interaction term has an ambiguity as the matrix $\mathbb{X}=\sqrt{g^{-1}f}$ is not completely determined, and a choice must be made. The solution chosen in this paper is the simplest one, and the one that can give continuous solutions in the presence of singularities \cite{Gratia:2013gka,Gratia:2013uza}.}. Note that eq.~(\ref{DHg}) and eq.~(\ref{DHf}) are redundant, and can be obtained by taking the derivative of the Friedmann equations (\ref{Friedg}) and (\ref{Friedf}), respectively.

In addition, we will have the matter conservation equation:
\begin{equation}\label{MatterConservation}
\rho'=-3\mathcal{H}(\rho+p),
\end{equation}
and the following Bianchi constraint:
\begin{equation}\label{Bianchi}
\left(pa^2\alpha\beta -m^4a_g^2 Z\right)\left(X_g\mathcal{H}_f-X_f\mathcal{H}_g\right)=0,
\end{equation}
where we have defined $Z\equiv \beta_1+2\beta_2N+\beta_3 N^2$. As previously mentioned, this last equation plays an important role as it divides the possible solutions into two branches. Branch I will be defined by the condition $(pa^2\alpha\beta -m^4a_g^2 Z)=0$ and Branch II by the condition $(X_g\mathcal{H}_f-X_f\mathcal{H}_g)=0$. In what follows we analyse these two branches, which have qualitatively very different solutions. In order to have a clear picture of the physical solutions, and for ease of comparison, this analysis will be mainly done using the Friedmann equation for the space-time metric $g^{\text{eff}}_{\mu\nu}$, which can be derived from the background equations presented above. In addition, from now on, without loss of generality, we choose the time coordinate $\tau$ such that $X=1$.

%---------------------------------------------------------------------------------------------------------------------------------------

\subsection{Branch I}

This branch is defined by the following Bianchi constraint:
\begin{equation}\label{Branch1constraint0}
(pa^2\alpha\beta -m^4a_g^2 Z)=0.
\end{equation} 
Let us analyse this constraint when $p=w\rho$, with $w$ constant. If $w\not= 0$, the constraint can be rewritten as:
\begin{equation}\label{Branch1constraint}
\rho =\frac{m^4Z}{(\alpha+\beta N)^2 w\alpha\beta}.
\end{equation}
Notice that for $w\not=0$, there is a particular choice of parameters that eliminates the $N$-dependence in eq.~(\ref{Branch1constraint}): $\beta\beta_1=\alpha\beta_2 $ with $\beta^2\beta_1=\alpha^2\beta_3$. However, this choice will be avoided as it does not give a valid description of the universe at all times, as it implies $\rho=constant$ always. In any other situation, this equation gives a relation between the ratio $N$ and the energy density $\rho$ of the perfect fluid.

Using eq.~(\ref{Friedmann}) and the constraint given by eq.~(\ref{Branch1constraint}) we can find the Friedmann equation for the space-time metric in this branch in terms of $a$, $N$ and $\rho$. However, expressed in this way the Friedmann equation has a rather complicated expression, which we explicitly show in Appendix \ref{App:Friedmann}, eq.~(\ref{FinalFriedmann}), together with its derivation. 
This Friedmann equation shows that, if some of the parameters $\beta_n$, $\alpha$ or $\beta$ had opposite signs, in general there would be a divergence in $\mathcal{H}$ at some finite time during the evolution of the universe, or a violation of positivity of one of the Friedmann equations (\ref{Friedg})-(\ref{Friedf})\footnote{Note that there might be very particular choices of parameters that avoid these two problems, but we do not explore these cases further as these choices would be very restricted.}. For this reason, we only consider solutions where all parameters have the same sign. Without loss of generality, from now on we assume all the parameters to be positive, in which case we will also assume $N\ge 0$, as a negative $N$ would violate the positivity of $\mathcal{H}_f^2$, rendering complex the solution for the physical metric. 

Assuming that all parameters are positive, and that $\alpha \not=0$ and $\beta \not=0$, we next analyse the possible solutions allowed by the constraint (\ref{Branch1constraint}) in two relevant regimes.

\begin{description}

\item[Early times:] We look for possible values of $N$ when $\rho/m^4\rightarrow \infty$. As we can see from the constraint (\ref{Branch1constraint}), if all parameters are positive, $\rho/m^4$ will never reach infinity, but a maximum value instead, which will depend on the choice of parameters. This maximum value can be found by taking the time derivative of eq.~(\ref{Branch1constraint}):
\begin{equation}\label{EqNb}
\rho' = \left(\frac{2m^4}{ w\alpha\beta}\right)\frac{\left[(\alpha\beta_2-\beta\beta_1)+N(\alpha\beta_3-\beta\beta_2)\right]}{(\alpha+\beta N)^3}N'\equiv F(N)N'.
\end{equation}
We see that the maximum will be reached when $F(N)=0$ or $N'=0$. For the case $\beta_3=0$ (which will be required later), it is possible to show that $N'>0$, except in an asymptotic limit, which actually leads to a minimum of $\rho$ given by $\rho=0$, as we will corroborate in the next section. Then, the maximum of $\rho$ will be determined by $F(N)$, and it will occur when $F(N_b)=0$, where:
\begin{equation}\label{EqNb2}
N_b	\equiv \frac{(\alpha\beta_2-\beta\beta_1)}{\beta\beta_2},
\end{equation}
where we have set $\beta_3=0$. The maximum of the energy density can be found by evaluating the constraint (\ref{Branch1constraint}) in $N=N_b$:
\begin{equation}\label{rhomaxb2}
\rho_{\text{max}}=\frac{m^4\beta_2^2}{w\alpha\beta^2\left(2\alpha\beta_2-\beta\beta_1\right)}.
\end{equation}
 
Notice that since $N'>0$ and $N\ge 0$, early times will in general be described by the period where $N \ll 1$, which does not necessarily correspond to the region around $N_b$, and thus around the maximum $\rho_\text{max}$. Therefore, if $N_b\sim 1$ or $N_b\gg 1$, we say that the maximum $\rho_\text{max}$ is reached at an intermediate stage, instead of early times.

The form that the Friedmann equation takes at early times and around $N_b$ will depend on the choice of parameters. As such, in the next section, we will show the evolution of the scale factor in detail in three different cases: $N_b>0$, $N_b=0$, and $N_b<0$.
Finally, we notice that since $N\ge 0$ we expect to have a maximum of the energy density at $N=N_b$ only if $N_b \ge 0$. If $N_b<0$, the value $N_b$ will never be reached, as the universe would start at $N=0$ because a negative $N$ would violate the positivity of $\mathcal{H}_f^2$. As we will see in the next section, in this last case, the maximum value of the energy density will be reached when $N=0$ and $\rho'(N=0)\not=0$.

\item[Late times:] We look for possible values of $N$ when $\rho/m^4\rightarrow 0$. From the constraint (\ref{Branch1constraint}) we see that the only possibilities are that $N\rightarrow 0$ (if $\beta_1=0$), or $N\rightarrow \infty$ (if $\beta_3=0$). Notice that these two cases are equivalent due to the transformation given by eq.~(\ref{Transf}). Therefore, from now on, without loss of generality, we will assume that $\beta_3=0$, and thus $N\rightarrow \infty$ when $\rho/m^4\rightarrow 0$, which will correspond to late-times. 

Notice that, if in addition we set $\beta_1=0$, we would also have $\rho/m^4\rightarrow 0$ when $N\rightarrow 0$. Since $N'>0$ and $N\ge 0$, the limit $N\rightarrow 0$ would correspond to early times, and thus if $\beta_1=\beta_3=0$, the universe will bounce, and $\rho/m^4\rightarrow 0$ in the infinite past and future.
 
If $\beta_4\not=0$, the Friedmann equation (\ref{FinalFriedmann}) at late times, i.e.~for $N\gg 1$, approximates to:
\begin{equation}\label{Branch1Late}
\mathcal{H} = H_0 a; \quad H_0 = m^2\sqrt{\frac{\beta_4}{3\beta^2}}, 
\end{equation}
where we recall that we have set $X=1$. Equation (\ref{Branch1Late}) gives a solution that corresponds to a de-Sitter phase, with a cosmological constant determined by the parameter $\beta_4$. 
\end{description}

%---------------------------------------------------------------------------------------------------------------------------------------

\subsection{Branch II}
Branch II has been previously studied in detail \cite{Enander:2014xga,Comelli:2015pua,Gumrukcuoglu:2015nua}, and here we simply summarise some relevant results found. 
This branch is defined by the following Bianchi constraint:
\begin{equation}\label{eqBranch1}
 X_g\mathcal{H}_f-X_f\mathcal{H}_g=0,
\end{equation}
and when combined with the two Friedmann equations (\ref{Friedg}) and (\ref{Friedf}), we get: 
\begin{align}
& m^4 \left[\beta_1+(3\beta_2- \beta_0)N + 3N^2(\beta_3- \beta_1)+ (\beta_4 -3\beta_2)N^3-N^4\beta_3\right]\nonumber \\
&+ (\beta -\alpha N)(\alpha+\beta N)^3\rho =0.\label{branchIIcondition}
\end{align}
Similarly to Branch I, eq.~(\ref{branchIIcondition}) gives a relation between the ratio $N$ and the energy density $\rho$ of the perfect fluid. 
Using eq.~(\ref{Friedmann}) and the constraint given by eq.~(\ref{branchIIcondition}), we find that the Friedmann equation for the space-time metric (the effective matter metric $g_{\rm eff}$) becomes: 
\begin{equation}\label{FriedBranchII}
\mathcal{H}^2= \frac{a^2}{3}\left[\alpha \rho(\alpha +N\beta)+ \frac{m^4}{(\alpha +N\beta)^2} (\beta_0+3N\beta_1+3N^2\beta_2+N^3\beta_3) \right].
\end{equation}

Assuming that $\alpha\not=0$ and $\beta\not=0$, we next analyse the solutions allowed by the constraint (\ref{branchIIcondition}) in two relevant regimes.

\begin{description}
\item[Early times:] We look for possible values of $N$ when $\rho/m^4\rightarrow \infty$. In this case, the constraint (\ref{branchIIcondition}) can be satisfied if $N\rightarrow \beta/\alpha$ or $N\rightarrow -\alpha/\beta$\footnote{Note that in both cases a particular relation between the parameters $\alpha$, $\beta$ and $\beta_n$ has to be assumed in order to satisfy the constraint in eq.~(\ref{branchIIcondition}).}.
As pointed out in \cite{Enander:2014xga}, the only viable situation is when $N\rightarrow \beta/\alpha$. In this case the Friedmann equation (\ref{FriedBranchII}) approximates to:
\begin{equation}
\mathcal{H}^2=\frac{ a^2}{3}(\alpha^2+\beta^2)\rho,
\end{equation}
which corresponds to GR with a modified gravitational constant. 

\item[Late times:] We look for possible values of $N$ when $\rho/m^4\rightarrow 0$. In this case, the constraint (\ref{branchIIcondition}) can be satisfied if $N\rightarrow 0$ (if $\beta_1=0$), $N\rightarrow \infty$ (if $\beta_3=0$), or $N\rightarrow \bar{N}$\footnote{Note that this particular constant value of $N$ is not arbitrary, but one such that the constraint (\ref{branchIIcondition}) is satisfied with $\rho=0$.}, where $\bar{N}$ is a non-zero constant. Notice that the two first types of solutions are related by the transformation given by eq.~(\ref{Transf}), and then they generate the same observables. Therefore, without loss of generality, we can only consider the case where $N\rightarrow \bar{N}$, where $\bar{N}$ is now a constant including zero.
Around the value $N=\bar{N}$, we find that the Friedmann equation (\ref{FriedBranchII}) approximates to:
\begin{equation}\label{FriedmLateB2}
\mathcal{H}^2= \frac{a^2}{3} \Lambda ,
\end{equation}
where $\Lambda$ is a constant given by:
\begin{equation}
\Lambda = \frac{m^4}{(\alpha +\bar{N}\beta)^2} (\beta_0+3\bar{N}\beta_1+3\bar{N}^2\beta_2+\bar{N}^3\beta_3).
\end{equation}

From eq.~(\ref{FriedmLateB2}) we see that the evolution of $a$ approaches a de-Sitter phase. In the particular case of $\bar{N}=0$, the cosmological constant will be given by the parameter $\beta_0$.

\end{description}

Particular choices of parameters were compared to data for Branch II in \cite{Enander:2014xga}, where viable (background) solutions for $a$ were described. However, at the level of linear perturbations tachyonic, gradient and ghost instabilities were found for tensor, vector and scalar perturbations, respectively \cite{Comelli:2015pua,Gumrukcuoglu:2015nua}, rendering this branch problematic, as perturbations would either break the perturbative approximation, or require fine-tuning.

%---------------------------------------------------------------------------------------------------------------------------------------
\section{Branch I}\label{Sec3:BranchI}

The presence of instabilities at the level of linear perturbations in Branch II motivates searching for viable cosmological solutions. In this section we address this problem, at the level of the background, by analysing the evolution of the space-time metric (\ref{sgeff}) in Branch I in detail. In particular, we distinguish three different choices of parameters, according to their values of $N_b$, and find approximate analytical solutions in some relevant regimes, as well as numerical solutions. As previously mentioned, we will be using the Friedmann equation for the physical metric given by eq.~(\ref{FinalFriedmann}) for analysing the behaviour of the scale factor $a$ in this branch, instead of the Friedmann equations for each metric $g_{\mu\nu}$ and $f_{\mu\nu}$. We do this for ease of comparison with other models, and as it is physically clear. As explained in Appendix \ref{App:Friedmann}, eq.~(\ref{FinalFriedmann}) can be written in terms of $N$ only, and therefore its behaviour can be simply analysed in terms of $N$, which is what we do throughout this section.
We recall that we will be assuming $\beta_3=0$, in order to have $\rho/m^4\rightarrow 0$ at late-times. 

%----------------------------------------------------------------------------------------------------------------------------
\subsection{Positive case: $N_b>0$ }

In order to have $N_b>0$, we could have $\beta_1=0$, or $\beta_1\not=0$ and $\beta_2\not=0$ such that $(\alpha\beta_2-\beta\beta_1)>0$. In what follows, we find the evolution of the scale factor $a$ during the radiation-dominated era, in three relevant regimes: early times ($N\ll 1$), intermediate bouncing regime ($N\sim N_b$), and late times ($N\gg 1$). 

\begin{itemize}
\item Assuming $w=1/3$, during early times, i.e.~for $N\ll 1$, eq.~(\ref{FinalFriedmann}) approximates to:
\begin{equation}\label{FriedEarly1}
\mathcal{H} = -\bar{H} a;\quad \bar{H} = m^2\frac{(\alpha\beta_2-\beta\beta_1)}{(\alpha\beta_2+\beta\beta_1)}\frac{\sqrt{\beta\beta_0+3\alpha\beta_1}}{\alpha\sqrt{3\beta}}>0,
\end{equation}
and if $\beta_1$ and $\beta_0$ are not zero at the same time, the evolution of the scale factor in conformal time will be given by $a=1/(\bar{H}\tau)$, which translates into the following solution in the physical time $t$:
\begin{equation}
a(t)=\bar{a}e^{-\bar{H} t}.
\end{equation}
Therefore, the universe is contracting, yet still accelerating with $\ddot a > 0$. Here, the value of $\bar{a}$ is determined by initial conditions. From eq.~(\ref{Branch1constraint}) we can also find the evolution on $N$ at early times:
\begin{equation}\label{Nbposrho0}
\rho\approx \frac{3m^4}{\alpha^4\beta}\left[\alpha\beta_1+2(\alpha\beta_2-\beta\beta_1)N\right],
\end{equation}
and then $N$ as a function of physical time is given by:
\begin{equation}\label{NEarlyNbpos}
N(t)=\frac{\alpha(\rho_{0r}\alpha^3\beta (m \bar{a})^{-4}e^{4\bar{H} t}-3\beta_1)}{6(\alpha\beta_2-\beta\beta_1)},
\end{equation}
where we have used that $\rho=\rho_{0r}/a^4$, where $\rho_{0r}$ is the energy-density of radiation today. As we can see from this equation, if $\beta_1\not=0$, as $t$ decreases, $N$ decreases and the value $N=0$ is reached at a finite time $t_0$, which gives a non-zero value of the scale factor $a(t_0)=\bar{a}e^{-\bar{H} t_0}$. The universe ends at this time, as otherwise $N$ would take negative values and thus $\mathcal{H}_f^2$ would become negative, rendering the solutions complex. 

On the other hand, if $\beta_1=0$, according to eq.~(\ref{NEarlyNbpos}), the value $N=0$ is reached in the infinite past $t\rightarrow -\infty$, limit at which $a\rightarrow \infty$. Thus, this case has an infinite contracting past, and no violation of positivity is present. For this reason, from now on we focus on the $\beta_2$-only case. We remark that when $t\rightarrow -\infty$, $N=0$ and $N'=0$. This asymptotic limit is the only one that gives $\rho'=0$ and $N'=0$. 

Finally, we analyse the next-to leading order term of eq.~(\ref{FriedEarly1}). When $\beta_1=0$, the Friedmann equation approximates to:
\begin{equation}
\mathcal{H}=-\bar{H}a+CNa; \quad C=\frac{1}{3}\frac{\sqrt{3}m^2(4\beta^2 \beta_0\beta_2^{1/2}-3\beta_2^{3/2} \alpha^2+\beta^2 \beta_0^{3/2})}{\sqrt{\beta_2\beta_0}\alpha^2\beta}.
\end{equation}
Since from eq.~(\ref{Nbposrho0}) we see that $\rho\propto N$ if we choose parameters such that $C<0$, the evolution of the universe would mimic GR with a cosmological constant (determined by $\beta_0$) at early times, but in a contracting universe. If we set $\beta_0=0$, the universe would approach GR without a cosmological constant (in a contracting universe, as before), as the Friedmann equation would approximate to:
\begin{equation}
\mathcal{H} =-m^2\sqrt{\frac{2\beta_2}{\alpha\beta}} \sqrt{N}a =-\alpha \sqrt{\frac{\rho}{3}}a,
\end{equation}
where in the last step we used eq.~(\ref{Nbposrho0}).

\item In what follows, we analyse the evolution of the scale factor in the intermediate regime given by $N\sim N_b=\alpha/\beta$. As previously explained, at this time the universe reaches a maximum of $\rho$, and therefore it bounces. Assuming $w=1/3$, for $N\sim N_b$, the Friedmann eq.~(\ref{FinalFriedmann}) approximates to:
\begin{equation}\label{eqHb1}
\mathcal{H} = B a(N-N_b); \quad B = \frac{\hat{H}_{gb}\hat{H}_{fg}\beta}{4\alpha(\beta \hat{H}_{gb}+\alpha \hat{H}_{fb})},
\end{equation}
where we have defined $ \hat{H}_{g}$ and $\hat{H}_{f}$ such that $X_g^2 a_g^2 \hat{H}_{g}^2=\mathcal{H}_g^2$ and $X_f^2 a_f^2 \hat{H}_{f}^2=\mathcal{H}_f^2$, and the subindex $b$ denotes evaluation at the bouncing time. In order to find the solution for $a$, we rewrite eq.~(\ref{eqHb1}) in terms of $a$ only. From the constraint (\ref{Branch1constraint}) we find that the relation between $a$ and $N$ around the bounce is given by:
\begin{equation}\label{RelaN1}
a=a_0+a_1(N-N_b)^2; \quad a_0=\left[\frac{2\rho_{0r}\alpha^2\beta^2}{3\beta_2 m^4}\right]^{1/4}, \quad a_1=\frac{a_0\beta^2}{16\alpha^2},
\end{equation}
where $\rho_{0r}$ is the energy-density of radiation today. Combining this last equation with eq.~(\ref{eqHb1}), the Friedmann equation becomes: 
\begin{equation}
\mathcal{H}= Ba_0\sqrt{\frac{(a-a_0)}{a_1}},
\end{equation}
whose solution in physical time is given by:
\begin{equation}
a(t)=a_0+\frac{B^2 a_0^2}{4a_1} (t-t_0)^2.
\end{equation}
As expected, there is a quadratic bounce in this regime, at which the scale factor takes its minimum value $a(t=t_0)=a_0$ given by eq.~(\ref{RelaN1}), which is completely determined by the parameters. At this minimum $a_0$, the energy density reaches a maximum given by eq.~(\ref{rhomaxb2}) with $\beta_1=0$:
 \begin{equation}\label{EqrhomaxNbpos}
\rho_{\text{max}}=\frac{3}{2}\frac{m^4\beta_2}{\beta^2\alpha^2}.
\end{equation}
Around the bounce, from eq.~(\ref{RelaN1}) we find that $N$ evolves as:
\begin{equation}
N(t)=N_b+\frac{Ba_0}{2a_1}(t-t_0).
\end{equation}
Finally, we remark that if $\beta_1\not=0$, the overall evolution for $a$ and $N$ during the bouncing period would have been the same.

\item Next, we analyse the evolution during late times, i.e.~for $N\gg 1$. In eq.~(\ref{Branch1Late}) we already found that the universe approached a de-Sitter phase determined by $\beta_4$ at late times, and then
\begin{equation}
a(t)=\bar{a}e^{H_0t},
\end{equation}
where $\bar{a}$ is determined by initial conditions and $H_0$ is given in eq.~(\ref{Branch1Late}). From the constraint (\ref{Branch1constraint}) we find that:
\begin{equation}\label{RhoNLate1}
\rho\approx \frac{6m^4\beta_2}{\alpha\beta^3 N},
\end{equation}
which means that $N$ evolves as
 \begin{equation}
N(t)=\frac{6m^4\beta_2}{\alpha\beta^3\rho_{0r}}\bar{a}^4e^{4H_0t},
\end{equation}
where $\rho_{0r}$ is the energy-density of radiation today. 
 
Next, we analyse the next-to leading order term in the large $N$ approximation:
\begin{equation}
\mathcal{H}= m^2\sqrt{\frac{\beta_4}{3\beta^2}}a+\frac{C}{N}a;\quad C=-\frac{m^2[(4\beta_4\alpha^2-3\beta^2\beta_2)\sqrt{\beta_2}+\alpha^2\beta_4^{3/2}]}{\alpha\beta^2\sqrt{3\beta_4\beta_2}}.
\end{equation}
But by using eq.~(\ref{RhoNLate1}) the Friedmann equation can be rewritten as:
 \begin{equation}\label{HLate1}
\mathcal{H}= m^2\sqrt{\frac{\beta_4}{3\beta^2}}a+\frac{C \alpha\beta^3 }{6m^4\beta_2}\rho a.
\end{equation}
From here we see that if $C>0$, at late times the evolution approaches GR with a modified gravitational constant and a cosmological constant. Similarly to the evolution at very early times, if $\beta_4=0$, the solution mimics GR without a cosmological constant, as the Friedmann equation approximates to: 
\begin{equation}\label{EqLateB1Nbpos}
\mathcal{H}= m^2\sqrt{\frac{2\beta_2}{\alpha\beta}}\frac{a}{\sqrt{N}}=\beta \sqrt{\frac{\rho}{3}}a.
\end{equation}
For this reason, we will require $\beta_4\not=0$ to obtain a cosmological evolution with late-time acceleration.
Finally, we remark that if $\beta_1\not=0$, the overall evolution for $a$ and $N$ during late times would have been the same. 

\end{itemize}

\begin{figure}[h!]
\begin{center}
\scalebox{0.34}{\includegraphics{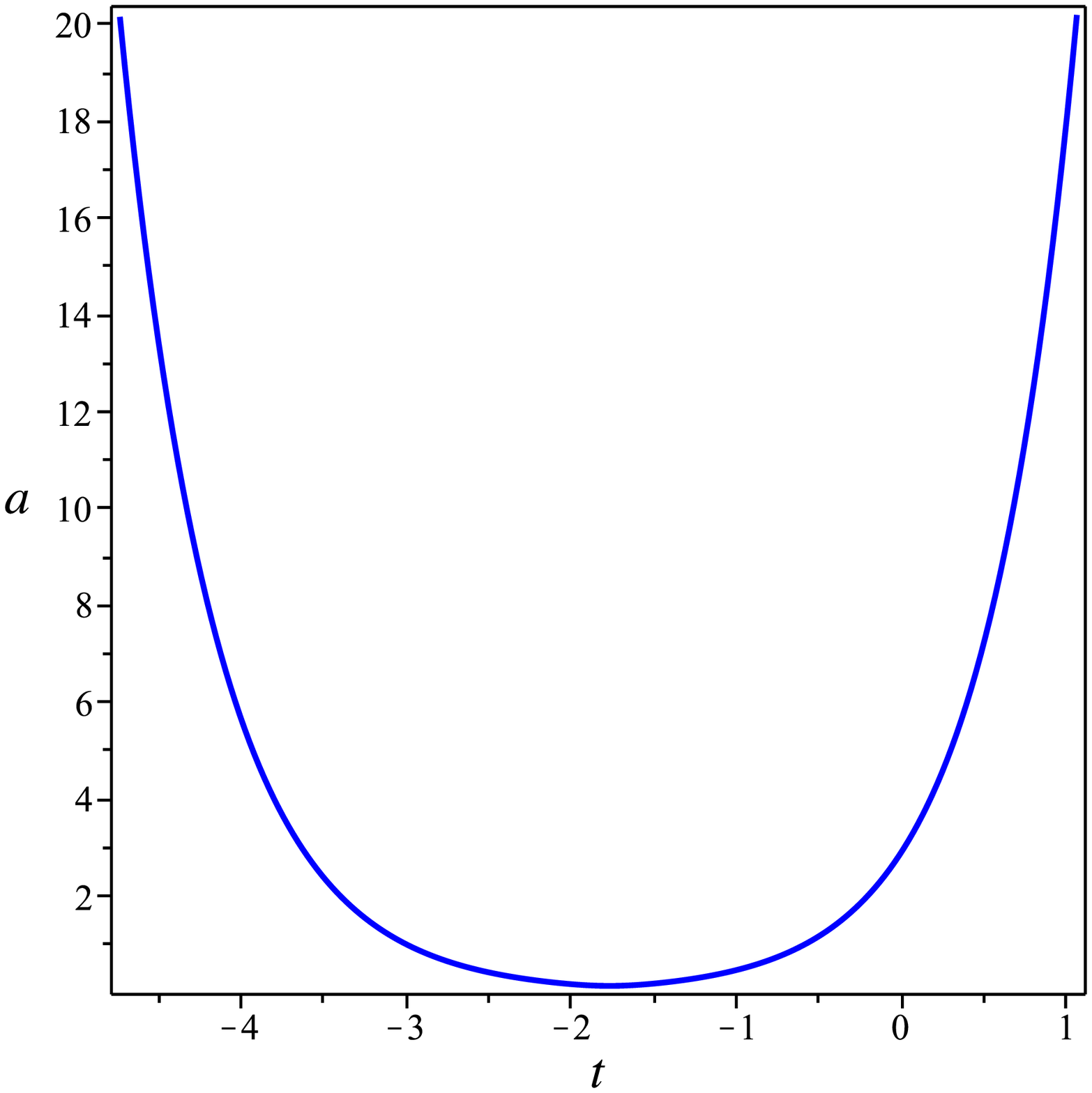}}
\scalebox{0.34}{\includegraphics{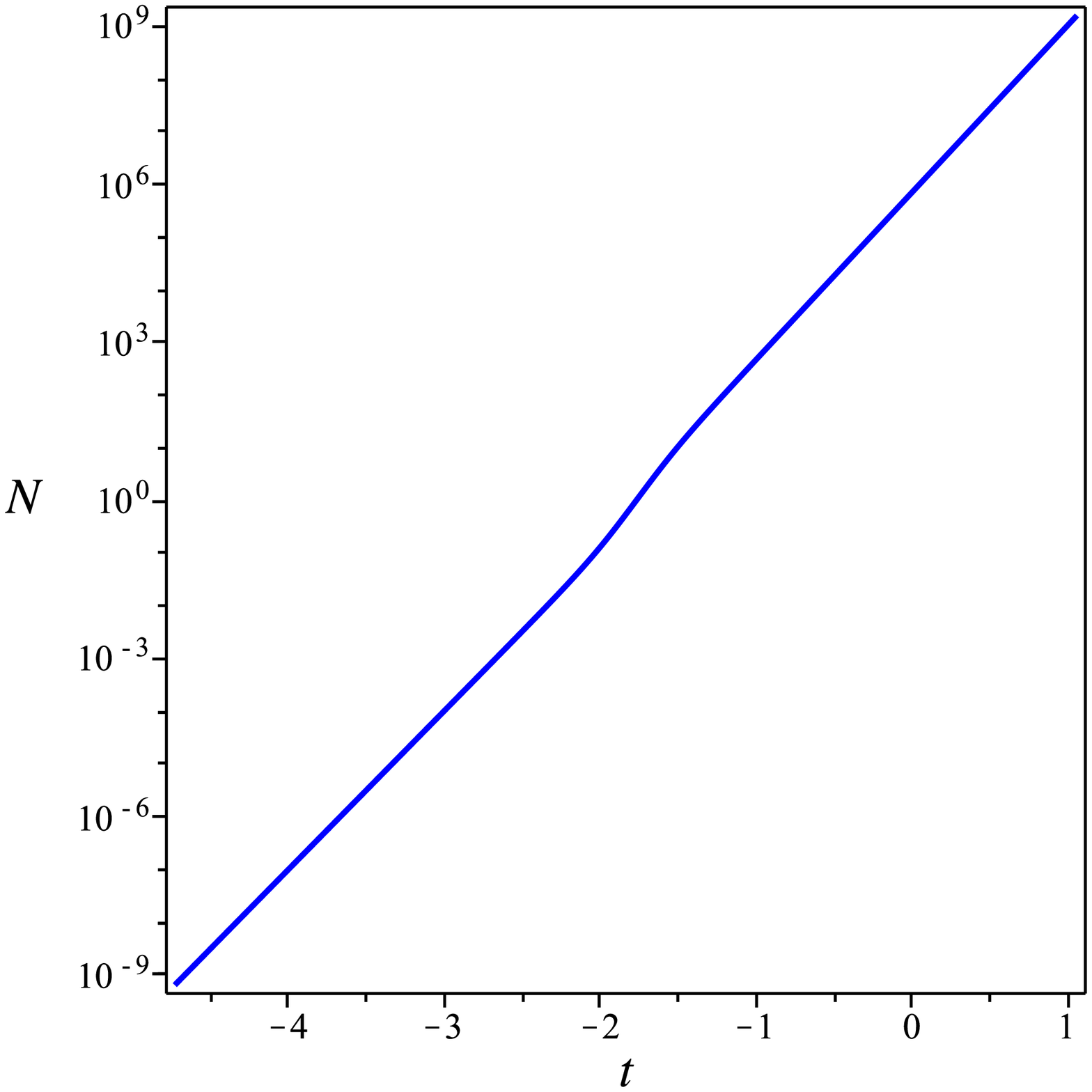}}
\caption{\label{fig:B2B4B0} Scale factor $a$ (LHS) and ratio $N$ (RHS) as a function of physical time $t$ during the radiation-dominated era for $\alpha=\beta=1$, $\beta_2=16$, and $\beta_4=10$, $\beta_0=9$, $m=1$. In this case $N_b>0$ and the bounce ocuurs when $N=1$.} 
\end{center}
\end{figure}

We now discuss numerical solutions for $a$, $N$, $a_g$, $a_f$ during the radiation-dominated era, as shown in in Figure \ref{fig:B2B4B0}. On the left-hand side of that figure, we show the scale factor $a$ as a function of physical time $t$, while on the right-hand side we show the ratio of scale factors $N$. We use arbitrary initial conditions and parameters such that $N_b>0$. 

\begin{figure}[h!]
\begin{center}
\scalebox{0.34}{\includegraphics{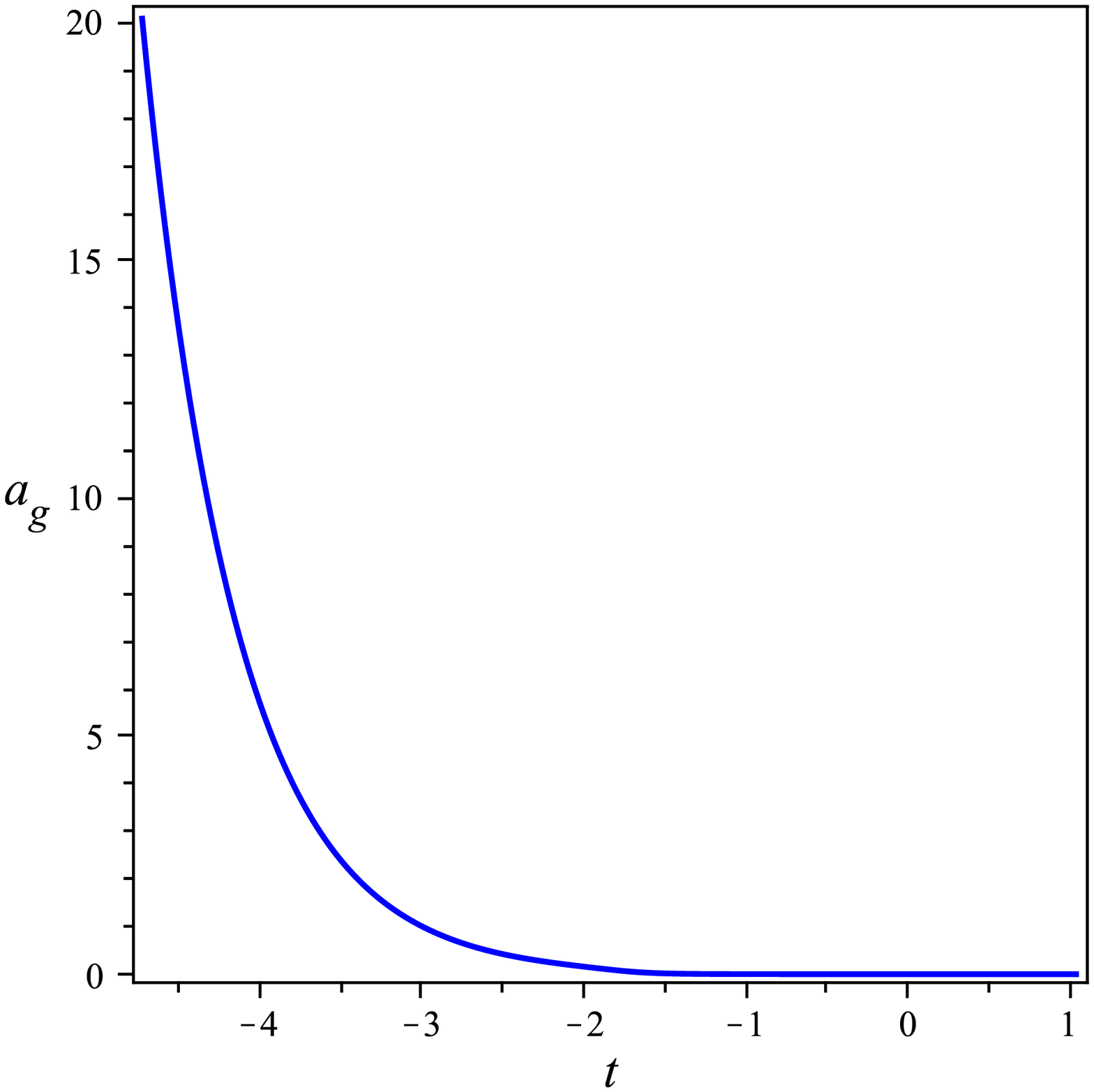}}
\scalebox{0.34}{\includegraphics{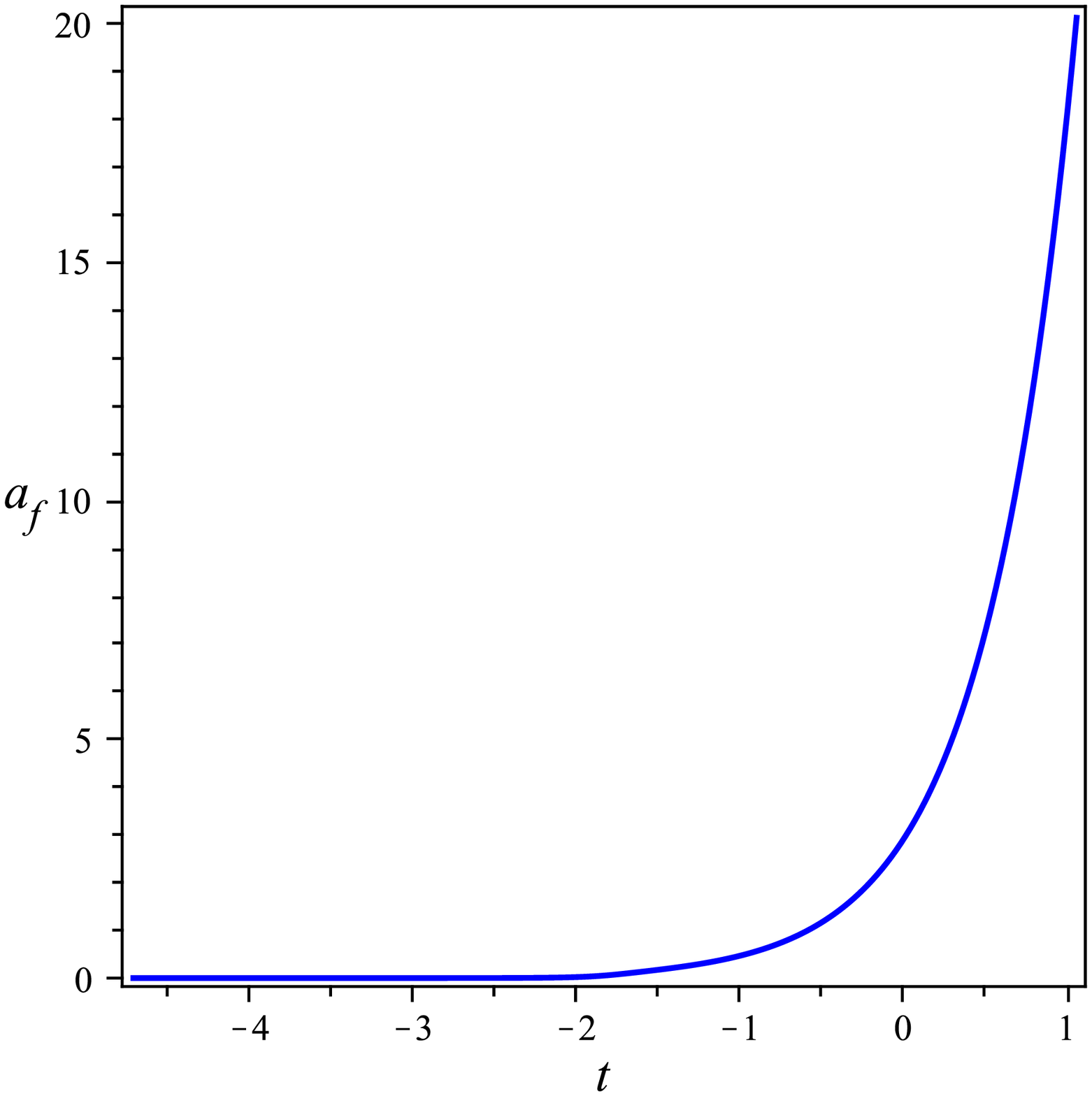}}
\caption{\label{fig:B2B4B0fg} Scale factor $a_g$ (LHS) and $a_f$ (RHS) as a function of physical time $t$ during the radiation-dominated era for $\alpha=\beta=1$, $\beta_2=20$, and $\beta_4=10$, $\beta_0=9$, $m=1$. In this case $N_b>0$.} 
\end{center}
\end{figure}

\FloatBarrier
As in the analytical solutions, in Figure \ref{fig:B2B4B0} we see that $N$ is a growing monotonic function, where $N \ll 1$ characterises early times, and $N\gg 1$ characterises late times. In addition, we observe the intermediate bouncing regime, with a minimum for the scale factor. In this numerical example we set $\beta_0\not=0$, which generates a contracting accelerated period for $a$ during early times, and $\beta_4\not=0$, which generates an expanding accelerated period at late times.
In Fig.~\ref{fig:B2B4B0fg} we show numerical solutions for the evolution of scale factors $a_g$ and $a_f$ as a function of physical time $t$ during the radiation-dominated era, for the same initial conditions and choice of parameters as in Fig.~\ref{fig:B2B4B0}. For this bouncing solution, we see that the scale factor $a_g$ always decays in time, whereas $a_f$ always grows in time. In fact, this behaviour can be derived from the previous analytical solutions found for $a$ and $N$. We find that at early times $a_g\propto e^{-\bar{H}t}$ and $a_f\propto e^{3\bar{H}t}$ (when $\beta_0\not=0$), and therefore $a_g\rightarrow \infty$ and $a_f\rightarrow 0$ in the infinite asymptotic past. During late times, $a_g\propto e^{-3H_0t}$ and $a_f\propto e^{H_0t}$  (when $\beta_4\not=0$), and thus $a_g\rightarrow 0$ and $a_f \rightarrow \infty$ in the infinite asymptotic future.

In summary, when $N_b>0$, the $\beta_1$ and $\beta_2$ case differs from the $\beta_2$-only case only during early times, as the former has a finite past with a non-zero value of the scale factor, whereas the latter has an infinite past with $a\rightarrow \infty$ for $t\rightarrow -\infty$. In the $\beta_2$-only case we found that the universe mimics GR with a cosmological constant at early times if $\beta_0\not=0$. In both cases there will be an intermediate bouncing regime with a non-zero minimum value of the scale factor, and a GR-like regime with cosmological constant at late times if $\beta_4\not=0$.
We also remark that, even though some of the solutions found did violate the positivity of $\mathcal{H}_f^2$ at some finite time, they could still be viable, as that time would represent the start of the evolution of the universe. This is different to what we would find if some of the parameters $\beta_n$, $\alpha$ and $\beta$ had different sign, where violations of positivity would be likely to happen at some time in the future, and hence the universe would end at a finite time in future. As previously mentioned, we discard this last kind of solutions as we restrain ourselves to only look for standard evolutions with an infinite future here.

%----------------------------------------------------------------------------------------------------------------------------
\subsection{Null case: $N_b=0$} In this case $\beta_1$ and $\beta_2$ need to be different from zero, and chosen such that $(\alpha\beta_2-\beta\beta_1)=0$. As previously explained, we expect a minimum of the scale factor $a$ (and a maximum of $\rho$) to be reached when $N=0$. 
Notice that, in this case, early times $N\ll 1$ coincide with the period $N\sim N_b=0$. Thus, there is no intermediate regime, and in what follows we analyse only two relevant regimes during the radiation-dominated era, namely early times and late times.

\begin{itemize}
\item At early times, i.e.~$N\ll 1$, the Friedmann equation (\ref{FinalFriedmann}) approximates to:
\begin{equation}\label{H1EarlyNb0}
\mathcal{H} = m^2\sqrt{\frac{\beta(3\beta_1\alpha+\beta\beta_0)}{12\alpha^4}}Na.
\end{equation}
In order to find the solution for $a$, we rewrite this equation in terms of $a$ only. From the constraint (\ref{Branch1constraint}) we find that:
\begin{equation}
\rho\approx \frac{3m^4\beta_1}{\alpha^3\beta}\left(1-\frac{\beta^2}{\alpha^2}N^2\right),
\end{equation}
and thus in this regime the explicit relation between $a$ and $N$ is given by:
\begin{equation}\label{RelaN2}
a=a_0+a_1N^2; \quad a_0=\left[\frac{\rho_{0r}\alpha^2\beta^2}{3\beta_2m^4}\right]^{1/4}, \quad a_1=\frac{\beta^2a_0}{4\alpha^2},
\end{equation}
where $\rho_{0r}$ is the energy-density of radiation today. By means of eq.~(\ref{RelaN2}), eq.~(\ref{H1EarlyNb0}) can be rewritten as:
\begin{equation}
\mathcal{H} = m^2\sqrt{\frac{\beta(3\beta_1\alpha+\beta\beta_0)}{12\alpha^4a_1}}a_{0}\sqrt{a-a_{0}}, 
\end{equation}
whose solution in physical time is given by:
\begin{equation}
a(t)=a_{0}+\frac{a_{0}m^4}{\beta}\frac{(3\beta_1\alpha+\beta\beta_0)}{12\alpha^2}(t-t_0)^2,
\end{equation}
where $t_0$ is determined by initial conditions. As expected, the early time period corresponds to the minimum $a$ period. Here, the scale factor evolves quadratically and reaches a minimum value $a_0$, given in eq.~(\ref{RelaN2}).

In addition, from eq.~(\ref{RelaN2}), we find the evolution for $N$ as a function of the physical time:
\begin{equation}
N(t)=\frac{m^2\sqrt{\beta(3\beta_1\alpha+\beta\beta_0)}}{\sqrt{3}\beta^2} (t-t_0),
\end{equation}
which means that the value $N=0$ is reached at a finite time $t_0$, which is the moment at which the evolution starts as $N$ cannot become negative (as before, this would violate the positivity of $\mathcal{H}_f^2$). In this case, the maximum value of the energy density will be given by eq.~(\ref{rhomaxb2}):
\begin{equation}
\rho_\text{max}=\frac{3m^4\beta_2}{\alpha^2\beta^2}.
\end{equation}

\item At late times, i.e.~$N\gg 1$, the solutions of $a$ and $N$ are the same as for the previous case with $N_b>0$, given by eq.~(\ref{HLate1}) if $\beta_4\not=0$, and by eq.~(\ref{EqLateB1Nbpos}) if $\beta_4=0$. Therefore, the late-time universe will have a cosmological constant when $\beta_4\not=0$. 
\end{itemize}

\begin{figure}[h!]
\begin{center}
\scalebox{0.34}{\includegraphics{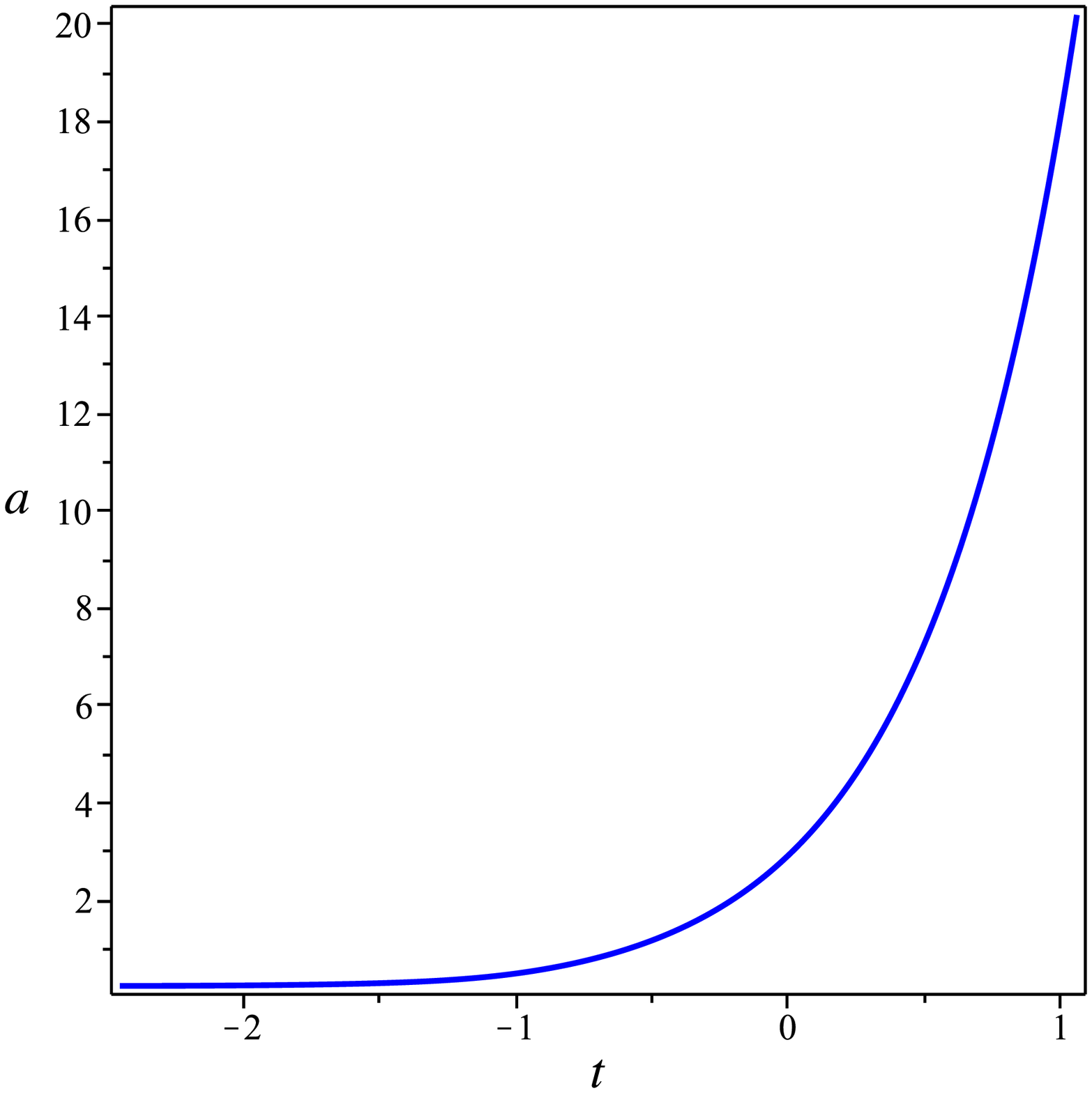}}
\scalebox{0.34}{\includegraphics{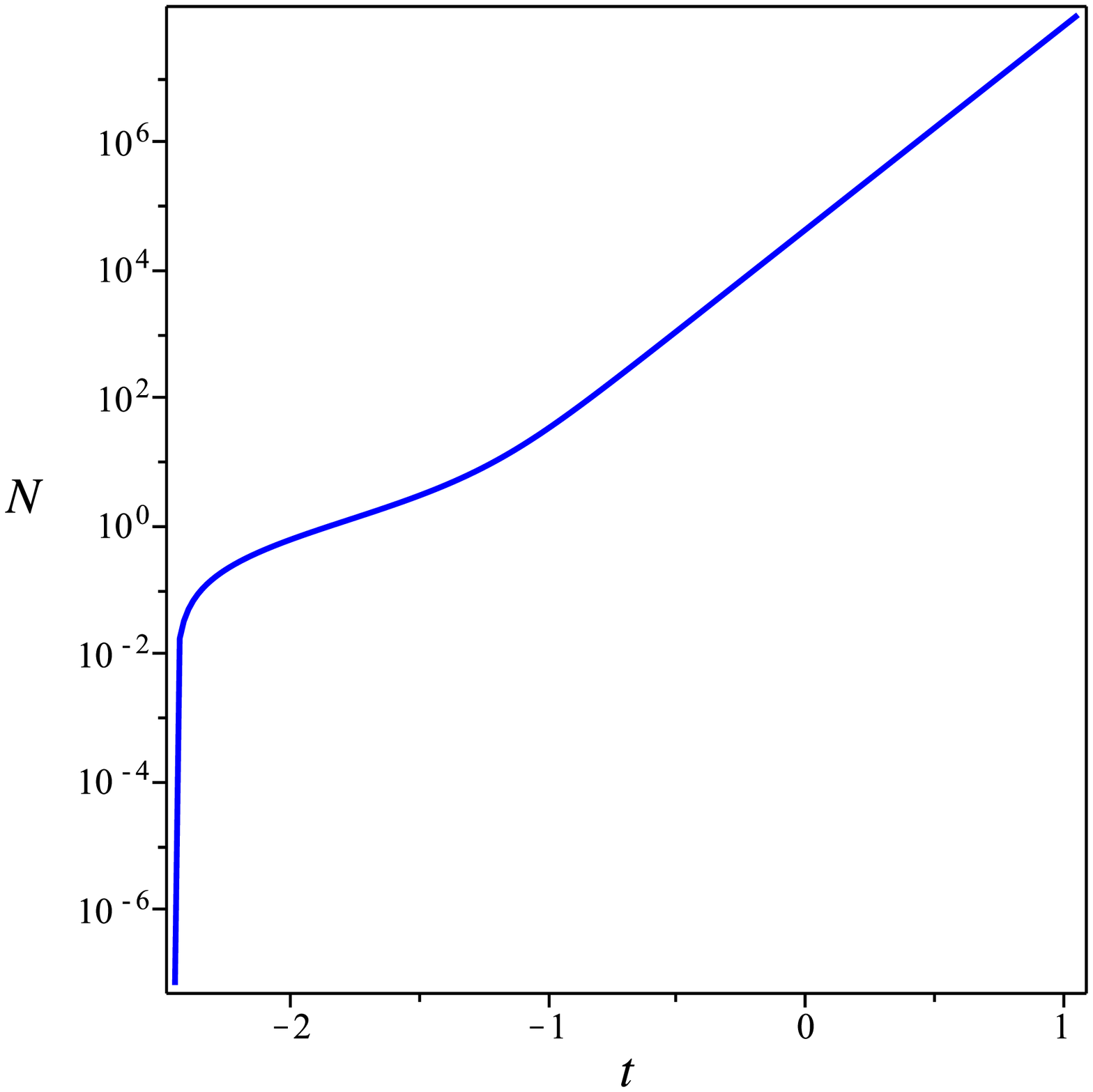}}
\caption{\label{fig:B1B2B4} Scale factor $a$ (LHS) and ratio $N$ (RHS) as a function of physical time $t$ during the radiation-dominated era for $\alpha=\beta=1$, $\beta_2=\beta_1=1$, and $\beta_4=10$, $\beta_0=0$, $m=1$. In this case $N_b=0$.} 
\end{center}
\end{figure}

Next, in Figure \ref{fig:B1B2B4}, we show numerical solutions during the radiation-dominated era and when $N_b=0$. On the left-hand side we show the scale factor as a function of physical time $t$, while on the right-hand side we show the ratio of scale factors $N$. We use arbitrary initial conditions and a choice of parameters such that $N_b=0$. As seen in the analytical solutions, $N$ is a growing monotonic function, where $N \ll 1$ characterises early times and $N\gg 1$ characterises late times. In addition, we have a finite past of the universe, where the scale factor $a$ starts at $t=-2.5$ at a non-zero minimum value. Since we set $\beta_4\not=0$ we observe a late-time accelerated expansion of the universe.

\begin{figure}[h!]
\begin{center}
\scalebox{0.34}{\includegraphics{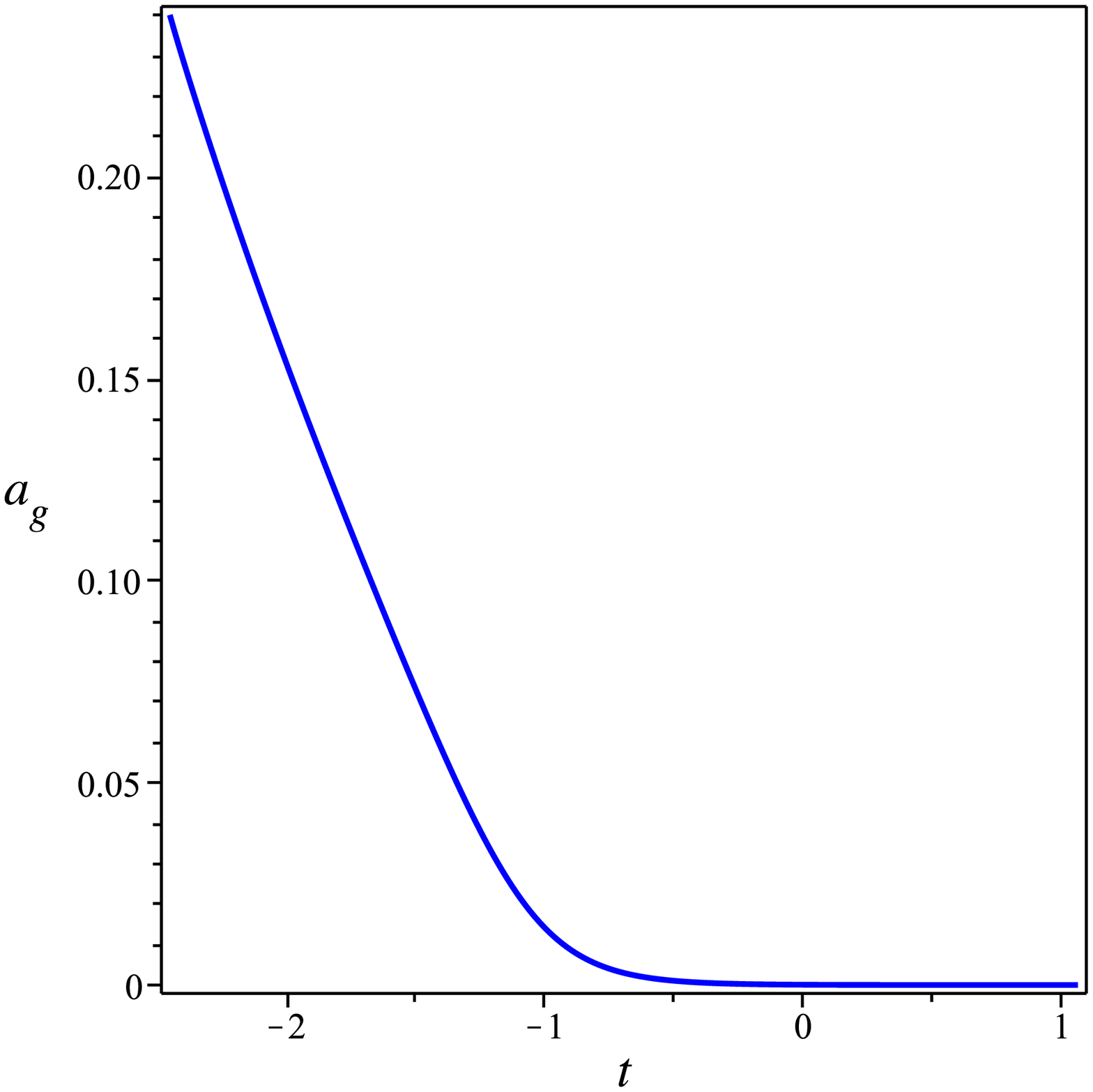}}
\scalebox{0.34}{\includegraphics{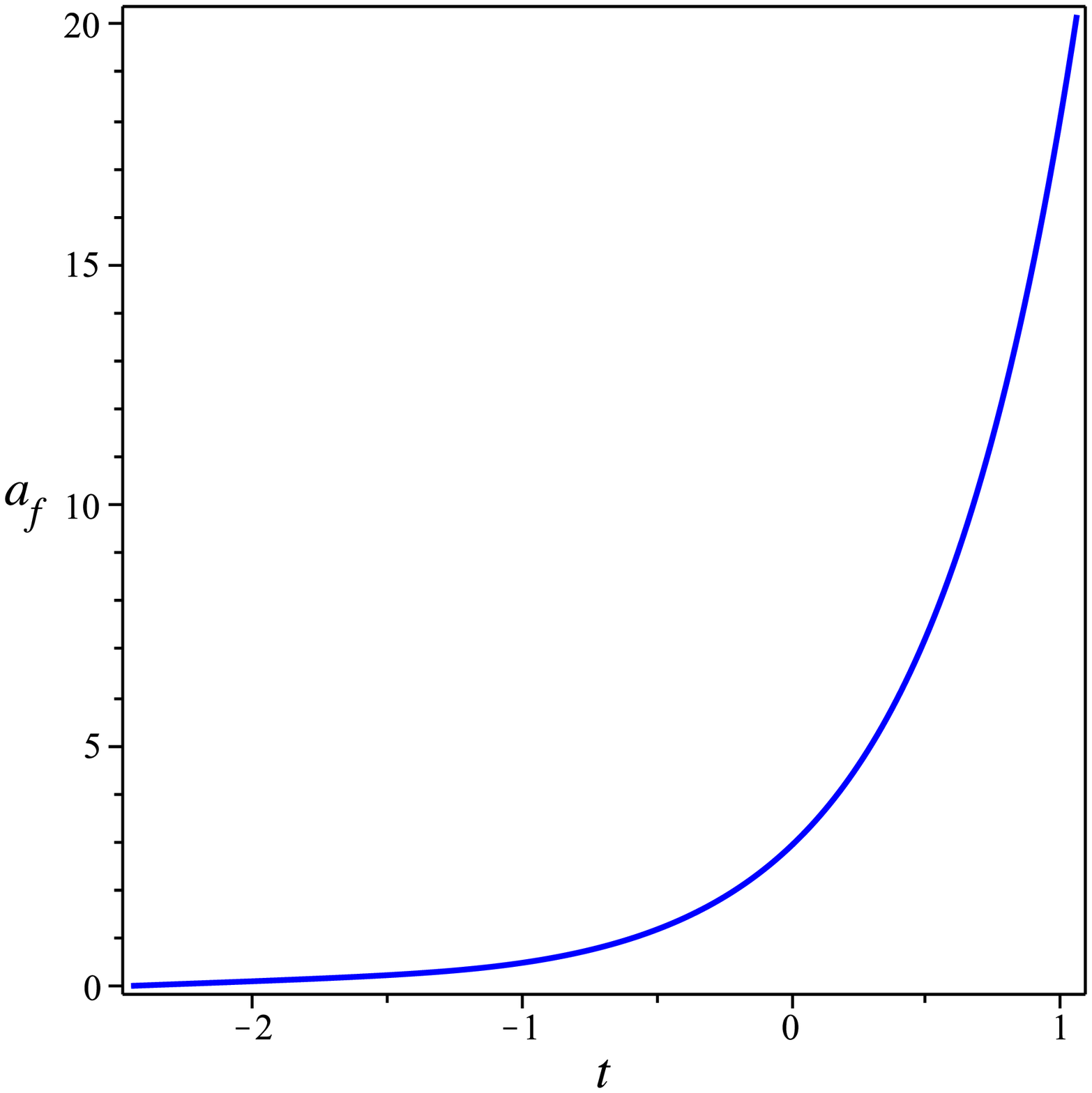}}
\caption{\label{fig:B1B2B4gf} Scale factor $a_g$ (LHS) and $a_f$ (RHS) as a function of the physical time during radiation-dominated era for $\alpha=\beta=1$, $\beta_2=\beta_1=1$, and $\beta_4=10$, $\beta_0=0$, $m=1$. In this case $N_b=0$.} 
\end{center}
\end{figure}

Furthermore, Figure \ref{fig:B1B2B4gf} shows the evolution of the scale factor $a_g$ and $a_f$ as a function of the physical time $t$ during the radiation-era. We chose the same initial conditions and parameter values as in Fig.~\ref{fig:B1B2B4}. In Fig.~\ref{fig:B1B2B4gf} we see that $a_g$ always decays in time, whereas $a_f$ always grows, similarly to the $N_b>0$ case. This behaviour can be derived from the previous analytical solutions found for $a$ and $N$. We find that during early times $a_g-a_{g\text{max}}\propto -(t-t_0)$ and $a_f\propto (t-t_0)$, where $a_{g\text{max}}$ is the maximum value of $a_g$ at the start of the evolution, at $t=t_0$. Then, at the beginning of the evolution $a_g=a_{g\text{max}}$ and $a_f=0$. On the contrary, during late-times $a_g\propto e^{-3H_0t}$ and $a_f\propto e^{H_0t}$ (if $\beta_4\not=0$), and therefore $a_g\rightarrow 0$ and $a_f \rightarrow \infty$ in the infinite future. 

Summarising, when $N_b=0$, we need $\beta_1\not=0$ and $\beta_2\not=0$ such that $(\alpha\beta_2-\beta\beta_1)=0$. The universe starts at a finite time $t_0$ at which there is a minimum non-zero value of the scale factor $a_0$, completely determined by the parameters (see eq.~(\ref{RelaN2})). The universe expands quadratically at early times, and approaches GR with a cosmological constant at late times when $\beta_4\not=0$.

%----------------------------------------------------------------------------------------------------------------------------
\subsection{Negative case: $N_b<0$} 

In this case we could have $\beta_2=0$, or $\beta_2\not=0$ and $\beta_1\not=0$ such that $(\alpha\beta_2-\beta\beta_1)<0$. As previously mentioned, there will be no bounce, as $N$ cannot be negative, and thus $N_b$ will never be reached. Thus, similarly to the previous case with $N_b=0$, we will analyse only two relevant regimes during the radiation-dominated era, namely early times and late times.

\begin{itemize}
\item During early times, i.e.~$N\ll 1$, the Friedmann equation is the same as eq.~(\ref{FriedEarly1}), but now $\bar{H}<0$. We can then write the Friedmann equation as:
\begin{equation}\label{FriedEarlyNbneg}
\mathcal{H} = |\bar{H}| a,
\end{equation}
and therefore the evolution of the scale factor in physical time is given by:
\begin{equation}
a(t)= a_0 e^{|\bar{H}| (t-t_0)},
\end{equation}
where $t_0$ is determined by initial conditions. We can see that during early times there is an exponential growth (regardless of the value of $\beta_0$). 
In what follows we find the evolution for $N$ also. From the constraint (\ref{Branch1constraint}) we find the relation between $a$ and $N$ to be:
\begin{equation}
a=a_0 + a_1 N; \quad a_0=\left[\frac{\rho_{0r}\alpha^3\beta}{3\beta_1m^4}\right]^{1/4}, \quad a_1=a_0\frac{(\beta\beta_1-\alpha\beta_2)}{2\alpha\beta_1}>0,
\end{equation}
where $\rho_{0r}$ is the energy-density of radiation today. Then, the evolution of $N$ in physical time is the given by:
\begin{equation}
N(t)= \frac{a_0}{a_1}\left(e^{|\bar{H}| (t-t_0)}-1 \right),
\end{equation}
%\begin{equation}
%N\approx \frac{\alpha\beta_1\left( 1-e^{-4|\bar{H}| t} \right)}{-2(\alpha\beta_2-\beta\beta_1)},
%\end{equation}
where we can see that $N$ grows exponentially in time, and $N=0$ is reached at a finite time $t_0$, when the scale factor $a$ reaches its minimum value $a_0$. At this time $t_0$ the universe starts as $N$ cannot be negative (otherwise it would violate the positivity of $\mathcal{H}_f^2$ as before). The maximum value for the energy density will given by eq.~(\ref{Branch1constraint}) evaluated at $N=0$:
\begin{equation}
\rho_{\text{max}}=\frac{3m^4\beta_1}{\alpha^3\beta}.
\end{equation}
Notice that, contrary to the previous cases, at $t=t_0$, $\rho$ is at its maximum, but $\rho'(t_0) \not=0$.
Finally, we mention that the next-to-leading order term in eq.~(\ref{FriedEarlyNbneg}) is proportional to $\sqrt{N}$. Since $N$ is related to $\rho$ at early times through eq.~(\ref{Nbposrho0}), this correction term goes as $\sqrt{\rho_{\text{max}}-\rho}$.

\item At late times, i.e.~$N \gg 1$, the Friedmann equation (\ref{FinalFriedmann}) approximates to eq.~(\ref{HLate1}) when $\beta_2\not=0$ and $\beta_4\not=0$, and eq.~(\ref{EqLateB1Nbpos}) when $\beta_2\not=0$ and $\beta_4=0$. 
On the other hand, if $\beta_2=0$, eq.~(\ref{FinalFriedmann}) approximates to:
\begin{equation}\label{FriedLateNbneg}
\mathcal{H} = m^2\sqrt{\frac{\beta_4}{3\beta}}a -\frac{1}{6}\frac{m^2\sqrt{2}\alpha\beta_4}{\beta^2 \sqrt{\beta_1}}\frac{a}{\sqrt{N}}.
\end{equation}
However, from the constraint (\ref{Branch1constraint}) we find that:
\begin{equation}
\rho\approx \frac{3m^4\beta_1}{\alpha\beta^3}\frac{1}{N^2},
\end{equation}
and then the next-to-leading order term in eq.~(\ref{FriedLateNbneg}) evolves as $\rho^{1/4}$. Therefore, if $\beta_2=0$, $a$ does not mimic GR, even when $\beta_4\not=0$. If in addition $\beta_4=0$, then
\begin{equation}
\mathcal{H}=m^2\sqrt{\frac{\beta_1}{\alpha\beta}}\frac{a}{N}=\beta \sqrt{\frac{\rho}{3}}a,
\end{equation}
and the solution asymptotically approaches GR without a cosmological constant at late times.

\end{itemize}

\begin{figure}[h!]
\begin{center}
\scalebox{0.34}{\includegraphics{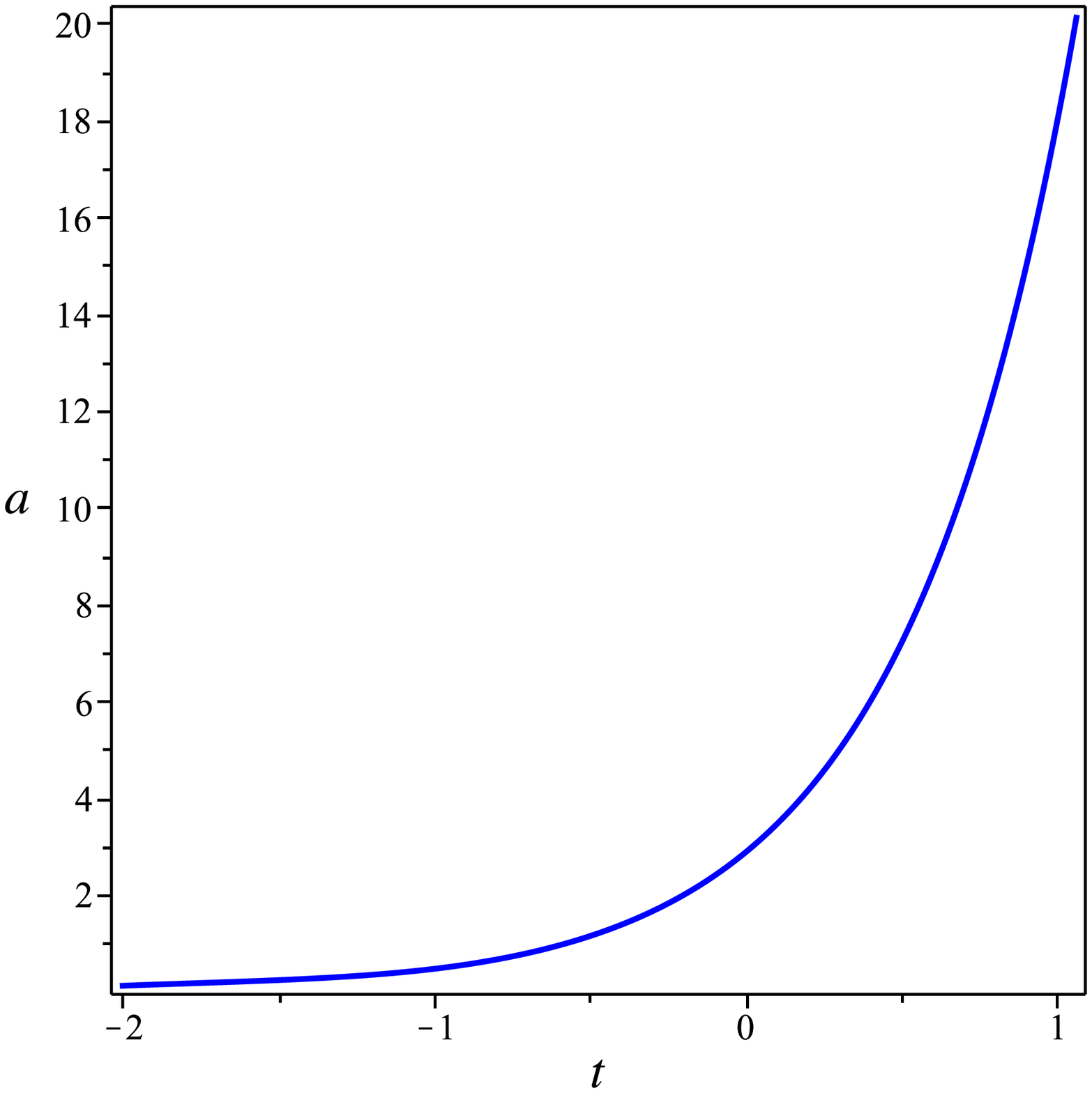}}
\scalebox{0.34}{\includegraphics{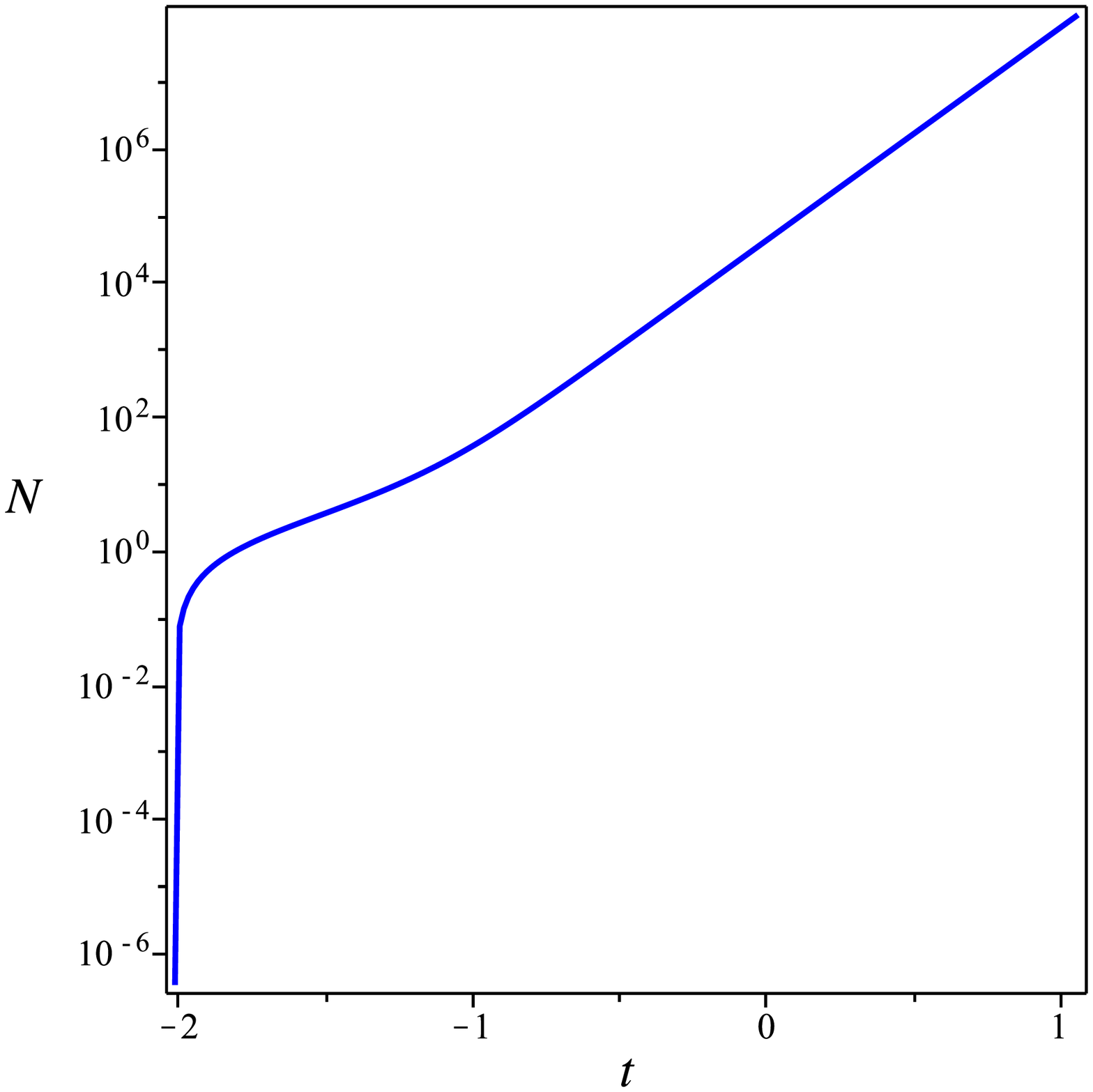}}
\caption{\label{fig:B1B2B4neg} Scale factor $a$ (LHS) and ratio $N$ (RHS) as a function of the physical time during radiation-dominated era for $\alpha=\beta=1$, $\beta_2=1$, $\beta_1=10$, and $\beta_4=10$, $\beta_0=10$, $m=1$. In this case $N_b<0$.} 
\end{center}
\end{figure}

A numerical solution during the radiation-dominated era when $N_b<0$ is shown in Figure \ref{fig:B1B2B4neg}. On the left-hand side we show the scale factor $a$ as a function of physical time $t$, while on the right-hand side we show the ratio of scale factors $N$. We use arbitrary initial conditions and parameters such that $N_b<0$. 

\begin{figure}[h!]
\begin{center}
\scalebox{0.34}{\includegraphics{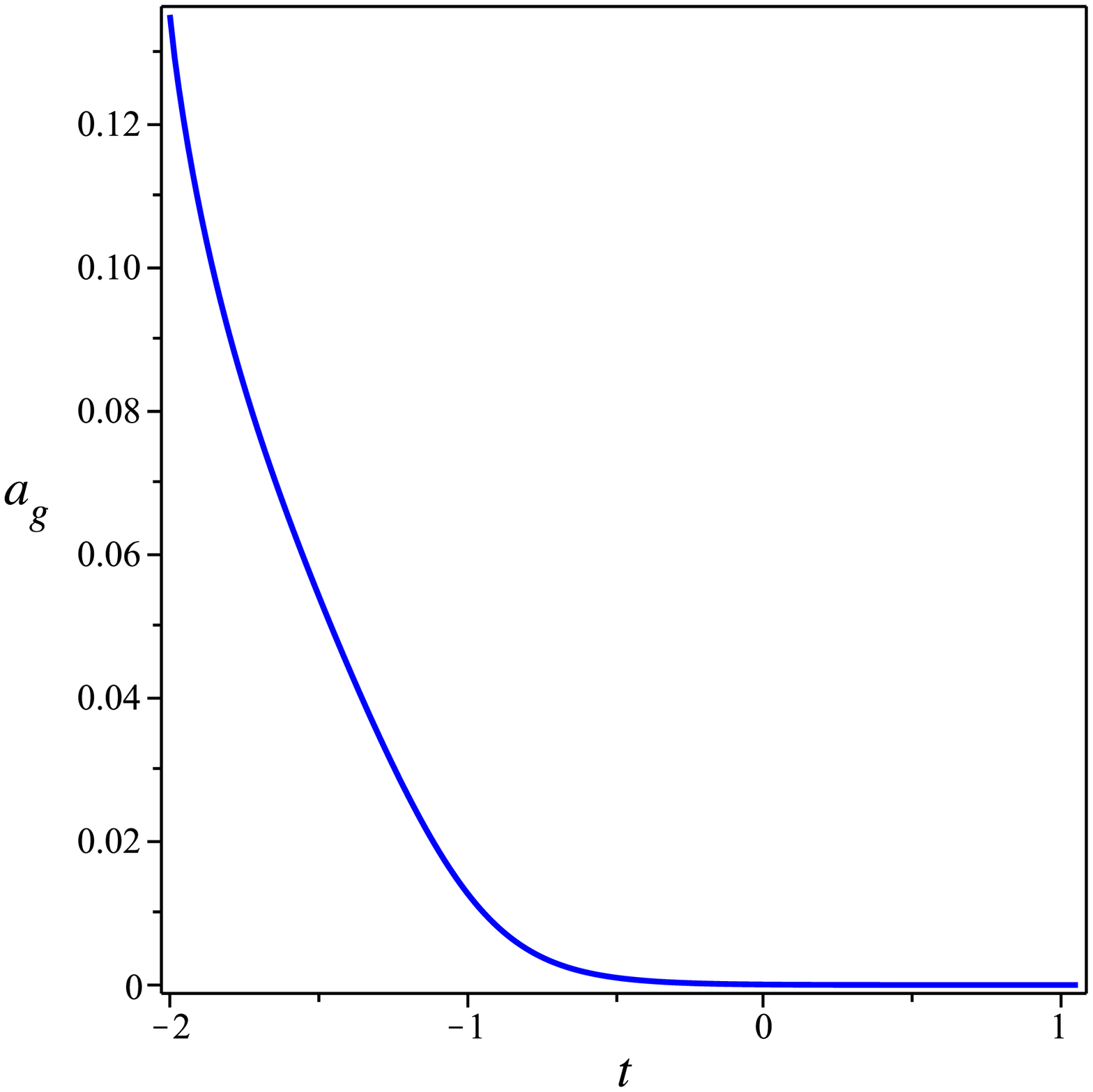}}
\scalebox{0.34}{\includegraphics{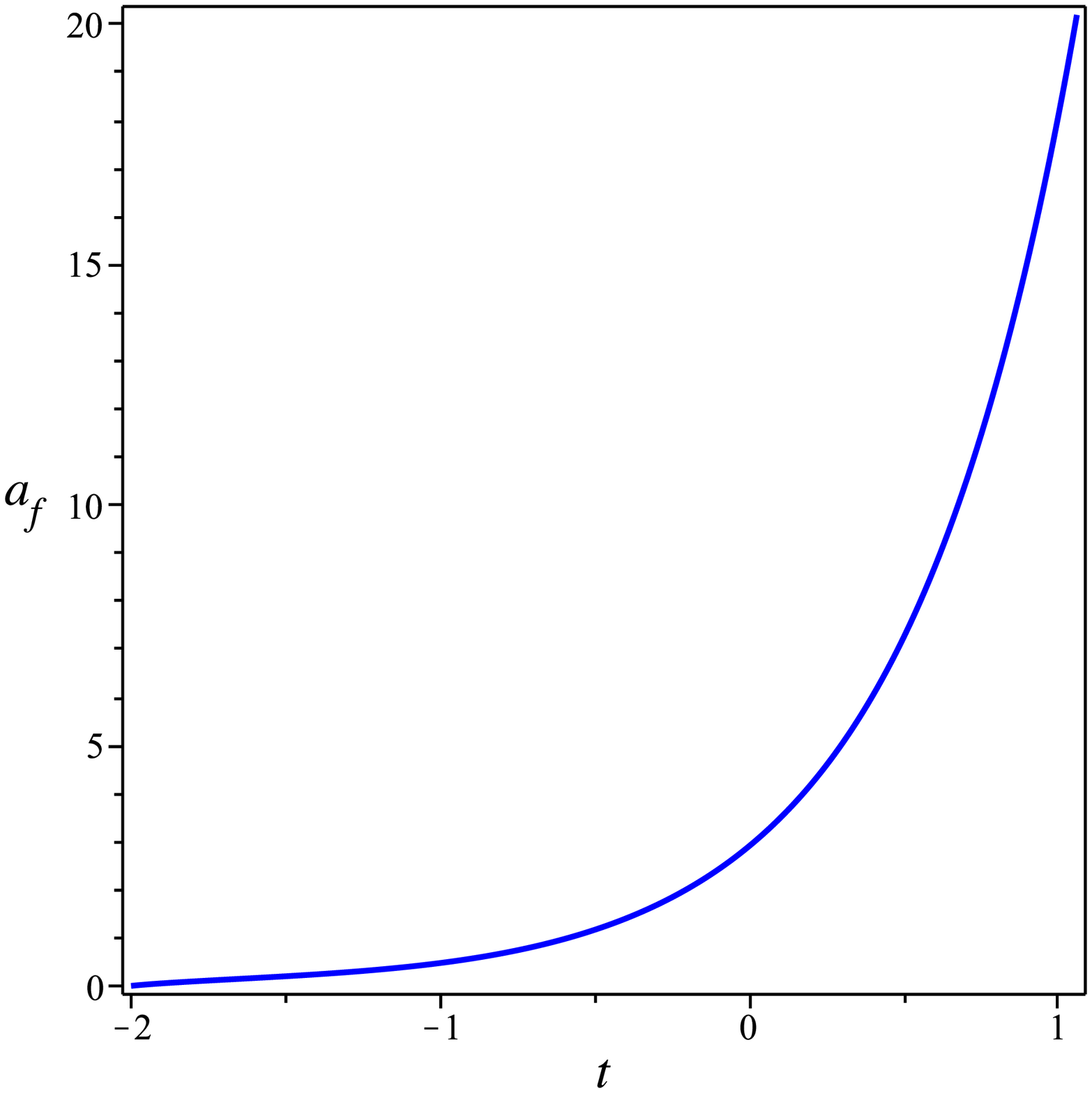}}
\caption{\label{fig:B1B2B4gfneg} Scale factor $a_g$ (LHS) and $a_f$ (RHS) as a function of the physical time during radiation-dominated era for $\alpha=\beta=1$, $\beta_2=1$, $\beta_1=10$, and $\beta_4=10$, $\beta_0=10$, $m=1$. In this case $N_b<0$.} 
\end{center}
\end{figure}

As seen in the analytical solutions, in Fig \ref{fig:B1B2B4neg} we see that $N$ is a growing monotonic function, where $N \ll 1$ characterises early times, and $N\gg 1$ characterises late times. In this solution the universe starts with a non-zero minimum value of the scale factor, with an accelerated early period. In addition, since we set $\beta_4\not=0$, we also have a late-time accelerated period. Furthermore, Figure \ref{fig:B1B2B4gfneg} shows the evolution of the scale factor $a_g$ and $a_f$ as a function of the physical time during the radiation-dominated era. We set the same initial conditions and parameter values as in Fig.~\ref{fig:B1B2B4neg}. In Fig.~\ref{fig:B1B2B4gfneg} we see that $a_g$ always decays in time, whereas $a_f$ always grows, similarly to the previous cases. In fact, this behaviour can be derived from the previous analytical solutions found for $a$ and $N$. We find that during early times the evolution is such that $(a_g-a_{g\text{max}})\propto (1-e^{|\bar{H}|(t-t_0)})$ and $a_f\propto (e^{|\bar{H}|(t-t_0)}-1)$, where at $t_0$ the universe starts. Thus, $a_g=a_{g\text{max}}$ and $a_f= 0$ at the beginning of the evolution. During late-times $a_g\sim e^{-3H_0t}$ and $a_f\sim e^{H_0t}$ (if $\beta_4\not=0$ and $\beta_2\not=0$), and therefore $a_g\rightarrow 0$ and $a_f \rightarrow \infty$ in the infinite future. 

Summarising, when $N_b<0$, the $\beta_1$ and $\beta_2$ case differs from the $\beta_1$-only case during late times only, as the former can mimic GR at those times, if $\beta_4\not=0$, whereas the latter cannot. During early times the evolution in both cases is the same, namely a finite past with a minimum scale factor $a_0$, and an exponential growth of $a$ in physical time.

%---------------------------------------------------------------------------------------------------------------------------------------------------
\subsection{Radiation and dust}

Previously, we studied the evolution of the universe in Branch I during the radiation-dominated era only. In what follows we include the effect of dust, focusing on the bouncing $\beta_2$-only solution. We emphasise that we do not assume radiation to be completely negligible as it is usually done.
This is because radiation plays a special role through the constraint given by eq.~(\ref{Branch1constraint0}), as this constraint depends solely on the pressure of the perfect fluid, hence radiation should not be fully neglected (see Appendix \ref{App:Dust}).
When adding dust, the Friedmann equation will be given by eq.~(\ref{FinalFriedmann}), where the explicit dependence on $\rho$ and $w$ will correspond to the values for radiation, while the implicit dependence on $\rho$ (via the functions $\hat{H}_g$ and $\hat{H}_f$ defined in the appendix), will correspond to the value for dust and radiation. This is due to the fact that in the derivation of eq.~(\ref{FinalFriedmann}) all its explicit dependence on matter came from the constraint (\ref{Branch1constraint0}), and therefore it only includes fluids with non-zero pressure. After making these replacements, we analyse the Friedmann equation in three relevant regimes, namely early times, bouncing period, and late times.

\begin{itemize}

\item At early times, i.e.~when $N \ll 1$, the Friedmann equation (\ref{FinalFriedmann}) approximates to:
\begin{equation}
\mathcal{H}= -m^2\sqrt{\frac{\beta_0}{3\alpha^2}}a - \frac{\alpha^3}{2m^2\sqrt{3\beta_0} } \rho_\text{m}a,
\end{equation}
where $\rho_\text{m}$ is the energy-density of dust. As we can see, the leading order term is the same as the one found during the radiation-dominated era in eq.~(\ref{FriedEarly1}) with $\beta_1=0$. However, the next-to-leading order term differs, but the solution still asymptotically approaches GR (in a contracting universe) at early times.
Similarly to the solution during the radiation-dominated era, if we set $\beta_0=0$, the Friedmann equation would be of the form:
\begin{equation}
\mathcal{H}=-\alpha \sqrt{\frac{\rho_\text{m}}{3}}a,
\end{equation}
and the solution would approach GR without a cosmological constant at early times.

\item Around the bounce, i.e.~when $N\sim N_b$, the Friedmann equation is given by eq.~(\ref{eqHb1}), where the only difference is that previously the explicit expressions for $\hat{H}_{gb}$ and $\hat{H}_{fb}$ in the coefficient $B$ included radiation only, but now they include radiation and dust. In addition, relation (\ref{RelaN1}) also holds when dust is included, and thus the expression for the minimum value of the scale factor does not change, and so it will depend on the radiation energy-density only.

\item During late times, i.e.~when $N \gg 1$, the Friedmann equation (\ref{FinalFriedmann}) approximates to:
\begin{equation}\label{HlateDust}
\mathcal{H}= m^2\sqrt{\frac{\beta_4}{3\beta^2}}a + \frac{\beta^3}{2m^2\sqrt{3\beta_4} } \rho_\text{m}a,
\end{equation}
where $\rho_\text{m}$ is the dust energy-density. Thus, the solution approximates GR with a cosmological constant at late times when $\beta_4\not=0$. 
Similarly to the results found during the radiation-dominated era, if $\beta_4=0$ the solution does not have a cosmological constant but still approaches GR (without a cosmological constant) at late times, as the Friedmann equation approximates to:
\begin{equation}\label{HlateDust2}
\mathcal{H}=\beta \sqrt{\frac{\rho_\text{m}}{3}}a.
\end{equation}

\end{itemize}

\begin{figure}[h!]
\begin{center}
\scalebox{0.34}{\includegraphics{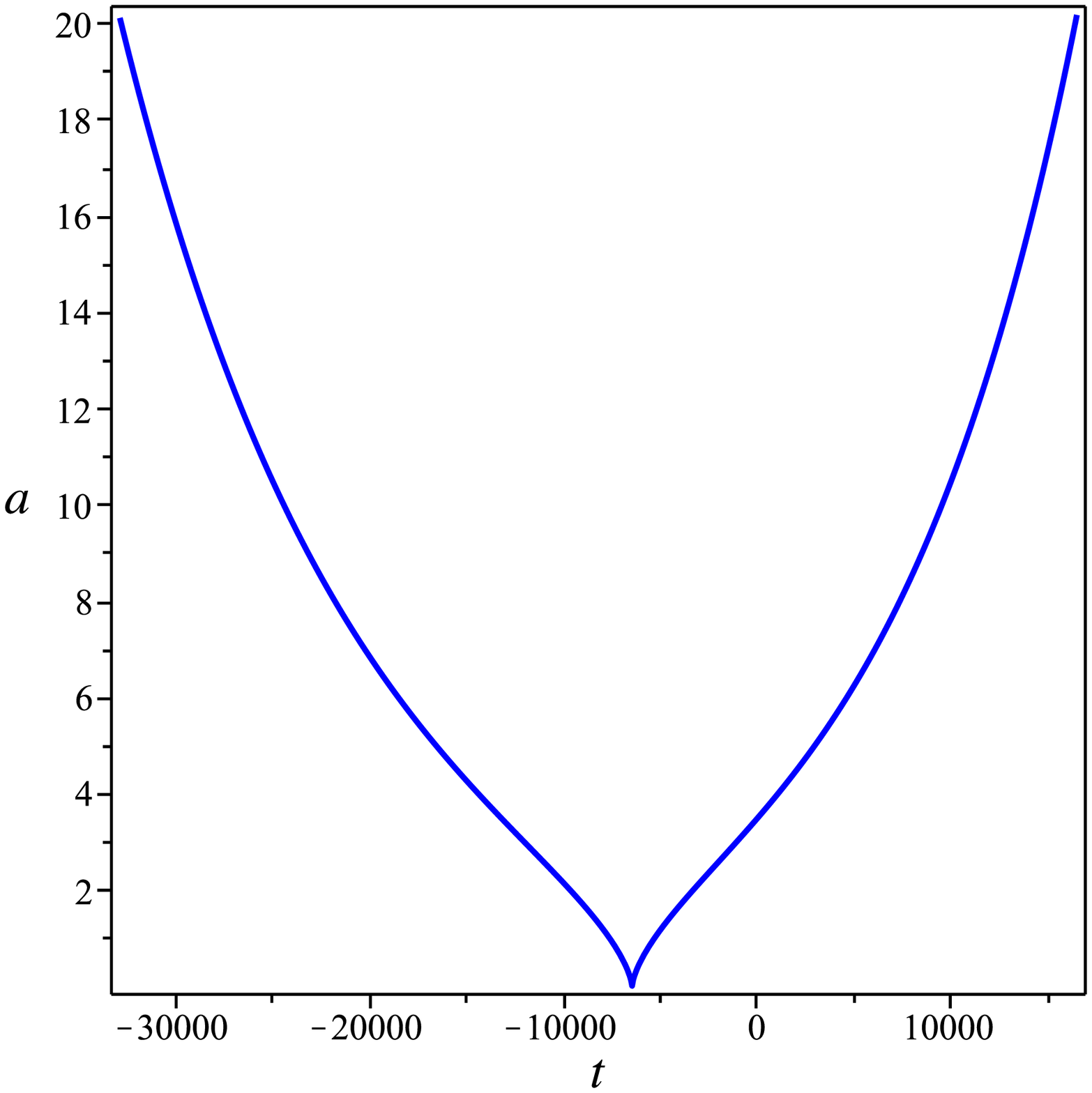}}
\scalebox{0.34}{\includegraphics{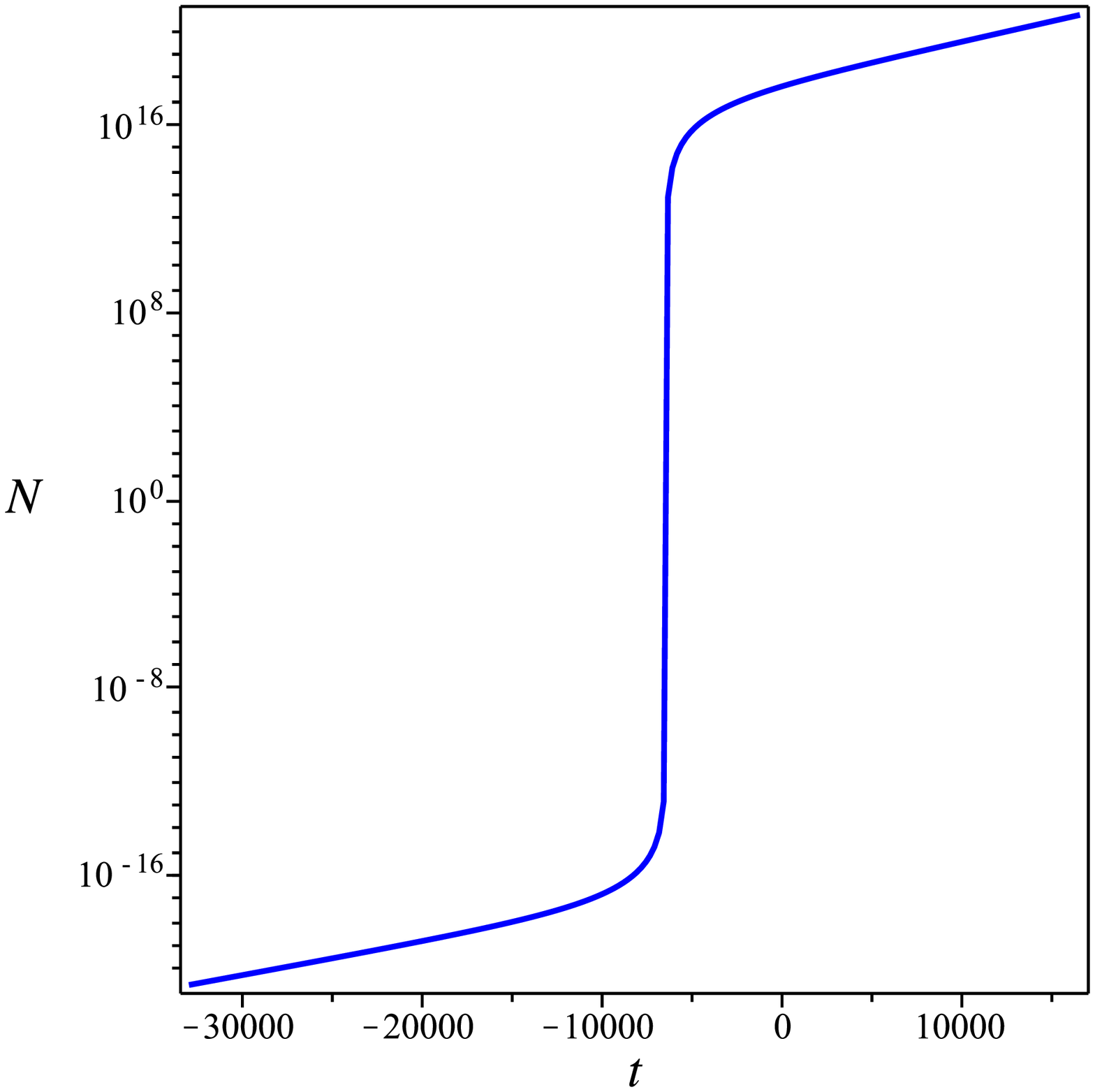}}
\caption{ \label{fig:BounceMatter} Scale factor $a$ (LHS) and $N$ (RHS) as a function of the physical time $t$ during radiation and matter dominated eras, when $\beta=\alpha=10^{-1}$, $\beta_4=3\times 10^{-10}$, $\beta_0=2\times 10^{-10}$, $\beta_2=5\times 10^4$, and $\beta_3=\beta_1=0$. In this case $N_b>0$.}
\end{center}
\end{figure}

Next, we show numerical solutions when radiation and dust are included. Figure \ref{fig:BounceMatter} on the left-hand side shows the evolution of the scale factor $a$ as a function of physical time $t$, while on the right-hide side it shows the evolution of the ratio of scale factors $N$. We use arbitrary initial conditions and a choice of parameters such that $N_b>0$. We observe the same overall behaviour as in Fig.~\ref{fig:B2B4B0fg}. Here we have chosen the parameters in a way that the evolution of the scale factor $a$ presents prolonged decelerating periods just before and after the bounce, which generate a difference between this figure and Fig.~\ref{fig:B2B4B0fg}. During these periods, the cosmological constants $\beta_0$ and $\beta_4$ do not yet dominate.

%---------------------------------------------------------------------------------------------------------------------------------------
\section{Summary, discussion and conclusions}\label{Sec:Conclusions}

In this paper we analysed a new branch of flat homogeneous and isotropic solutions in massive bigravity with double matter coupling. In this branch, even though generic choices of parameters can lead to divergences of the Hubble rate $\mathcal{H}$ at finite times and/or violations of positivity of the equations, the family of parameters $\beta_n,\alpha,\beta>0$ gives viable cosmological background solutions. All these solutions share the following characteristics: they have a minimum non-zero value of the scale factor, an accelerated period during the radiation-dominated era, and an accelerated period at late times if $\beta_4\not=0$. Furthermore, all the solutions have a finite past, except when $\beta_1=\beta_3=0$. This last case is the only one that avoids divergences and violations of positivity at all times, and thus gives a universe with an infinite past and future. 
In Table \ref{TableNbpos} we summarise the main solutions found in this branch.
\begin{table}[h!]
\centering
\begin{tabular}{| c || c || c || c || c |}
\hline
  & Non-zero $\beta_{1,2,3}$ & Past & Intermediate & Future \\ \hline \hline
 \multirow{2}{*}{$N_b>0$} & $\beta_1$-$\beta_2$ & {\footnotesize Finite exponential/polynomial contraction} & Bounce & GR+$\Lambda$\\ 
 & $\beta_2$ & {\footnotesize Infinite exponential/polynomial contraction} & Bounce & GR+$\Lambda$ \\ \hline
 $N_b=0$ & $\beta_1$-$\beta_2$ & Finite quadratic expansion & - & GR+$\Lambda$ \\ \hline
 \multirow{2}{*}{$N_b<0$} & $\beta_1$-$\beta_2$ & Finite exponential expansion & - & GR+$\Lambda$ \\ 
 & $\beta_1$ & Finite exponential expansion & - & not GR+$\Lambda$ \\ \hline
\end{tabular}
\caption{\label{TableNbpos} Summary of evolutions of the scale factor $a$ in Branch I when $\beta_n,\alpha,\beta>0$. In the first column, solutions were separated in three categories according to the value of $N_b$ (see eq.~(\ref{EqNb2})), namely when $N_b>0$, $N_b=0$, and $N_b<0$. We recall that $N_b$ corresponds to the value of the ratio of scale factors $N = a_f/a_g$ at the point where the energy density satisfies $\rho'=0$ (and reaches its maximum $\rho_{\rm max}$ for $N_b>0$, or $N_b=0$). In the second column we mention the non-zero interaction parameters ($\beta_1$, $\beta_2$, and $\beta_3$). Notice that it is always assumed that $\alpha\not=0$ and $\beta\not=0$. In the three cases we set $\beta_3=0$ as this gives an infinitely expanding universe in the future. In the rest of the table we describe the evolution of the universe in the past and future, and during an intermediate bouncing stage, if present. All solutions have a finite past except when $N_b>0$ and $\beta_2\not=0$, where there is an infinite contracting past. In addition, during the past, the contraction can be exponential ($\beta_0\not=0$ or $\beta_1\not=0$) or polynomial ($\beta_0=\beta_1=0$) as a function of the physical time $t$ when $N_b>0$. For $N_b=0$, there will always be a quadratic expansion, whereas for $N_b<0$ there will always be an exponential expansion. On the other hand, when $N_b>0$, solutions present a bouncing intermediate stage. Finally, at late times all solutions asymptotically approach GR with a cosmological constant if $\beta_4\not=0$, except when $N_b<0$ and $\beta_1\not=0$, where the solutions do not mimic GR but do have a cosmological constant.}
\end{table}

The solutions we find are interesting in both early and late universe settings. In an early universe context, they naturally lead to an early-time inflating period, without the assumption of a fundamental yet unknown field with very constrained self-interactions driving this early-time accelerated expansion. In addition, since the minimum value of the scale factor is not zero (and the energy density does not diverge), these types of solutions also remove the physical Big Bang singularity present in GR. All these features are new in massive bigravity with both single and double matter coupling, as all previously found viable cosmological solutions started at $a=0$ in an early decelerating epoch filled with radiation. 
However, it is important to notice that massive bigravity with double matter coupling is an effective field theory with an unknown cutoff, at least equal or higher than its strong coupling scale given by $\Lambda_3$. Therefore, for the solutions found to be viable alternatives to standard inflation, the parameters should be chosen in a way that the maximum $\rho_\text{max}$ happens at a lower energy scale than the cutoff scale of the theory, otherwise the inflating and/or bouncing feature cannot be trusted. For the usual choice of parameters, i.e.~$m\sim 10^{-33}$eV, the scale $\Lambda_3$ would be of the order of $10^{-13}$eV, and thus if this was the cutoff scale of the theory, no early time evolution found could be trusted. However, in \cite{Matas:2015qxa} an upper bound for the cutoff of the theory was found, given by $[(m^3M_\text{pl}^2)/(\alpha\beta)]^{1/5}$. If this was the actual cutoff, for instance for the bouncing solution with $\beta_1=\beta_3=0$, if $m\sim 10^{-33}$eV the bounce could in principle be within the regime of validity of the theory if it happened at an energy scale of around $10^7$eV\footnote{Here we have considered $\alpha\beta\sim 10^{-80}$ and $\beta_2\sim 1$ in order to get $\rho_\text{max}^{1/4}\sim 10^{7}$eV, according to eq.~(\ref{EqrhomaxNbpos}). For larger values of $\alpha\beta$ the cutoff and bouncing scales would be smaller than $10^7$eV, and thus BBN could not take place. For smaller values of $\alpha\beta$ the cutoff and bouncing scales would be larger than $10^7$eV, but the bouncing scale would be larger than the cutoff scale, so the bouncing feature would be outside the regime of validity of the effective field theory. This choice of $\alpha,\beta$ is of course fine-tuned -- an examination of the naturalness of the parameter values chosen here is beyond the scope of this paper, however.}. As this energy scale is approximately that of Big Bang nucleosynthesis, we would like the cutoff {\it and} the bouncing scale to be higher, which could be achieved by considering a larger $m$ and a larger $\alpha\beta$. However, the tuning on the parameter $m$ (and $\alpha,\beta$) must be done carefully, as it cannot be arbitrarily large, if massive bigravity is to satisfy solar system gravity constraints as well. Clearly, a detailed analysis of observational and technical naturalness constraints is required to know if (and in what sense) solutions in Branch I could be alternatives to standard inflation. Regardless of whether this is the case or not, we emphasise that these solutions are of relevance in a late-universe context. As we have shown throughout this paper, they provide consistent cosmological evolutions at those late times, closely mimicking GR with a cosmological constant in most cases. 

Further work is necessary to determine the full viability of Branch I solutions beyond the background behaviour studied in this paper. Here we have analysed their main properties, but a detailed study contrasting the solutions we found with data should be carried out. Furthermore, it is necessary to analyse the stability and evolution of cosmological perturbations around these backgrounds. A general analysis of linear perturbations in Branch I and II was previously performed in \cite{Gumrukcuoglu:2015nua}. Following the standard SVT decomposition \cite{Mukhanov:1990me} and the results of \cite{Gumrukcuoglu:2015nua}, in \cite{uspert} we argue that it is possible to avoid ghost and gradient instabilities on tensor and vector perturbations in Branch I, but tachyon instabilities will be present in both. In particular, for the bouncing solution with $\beta_1=\beta_3=0$, ghost and gradient instabilities can be avoided when $\beta_4>\beta_2$, $\beta_0>\beta_2$ and $\beta_4\alpha^2+15\beta^2\beta_2>\beta_0\beta^2+15\beta_2\alpha^2$. Due to the involved expressions describing scalar perturbations, we do not comment on their stability further here. Instead, we leave this issue for future work.

We finally remark that, within the bouncing solutions with $\beta_1=\beta_3=0$, we might find partially massless bigravity theories \cite{Deser:1983mm,Hassan:2012gz},
defined by the following choice of parameters:
\begin{equation}\label{EqBPartial}
\beta_0=3\beta_2=\beta_4, \quad \beta_1=\beta_3=0.
\end{equation}
Arguably the most interesting property of these theories is that around a de-Sitter space (both metrics are de-Sitter\footnote{This is different to the late-time de-Sitter phase found in Branch I, where the effective metric expanded exponentially in time, but the individual metrics $f_{\mu\nu}$ and $g_{\mu\nu}$ did not. In fact, we recall that at late-times the individual scale factors evolved as $a_g\propto e^{-3H_0t}$ and $a_f\propto e^{H_0t}$.}) they display an extra gauge symmetry (besides diffeomorphism invariance) that removes the helicity-0 mode of the massive graviton at linear order (and beyond, should it turn out to be a full gauge symmetry). Due to the extra symmetry, these theories propagate 6 DoF, as opposed to the 7 DoF propagated in massive bigravity. If this gauge symmetry can be non-linearly completed, these models would provide a very interesting modification to GR.
It is not known if this gauge symmetry appears around the backgrounds found in Branch I, but this should become clear when studying linear perturbations around such backgrounds in the future.

%---------------------------------------------------------------------------------------------------------------------------------------

\begin{acknowledgments}

We are grateful to Pedro Ferreira for useful discussions and comments. ML was funded by Becas Chile. JN acknowledges support from the Royal Commission for the Exhibition of 1851 and BIPAC.

\end{acknowledgments}

%---------------------------------------------------------------------------------------------------------------------------------------
\appendix

\section{Branch I: Friedmann equation for effective metric}\label{App:Friedmann}

In this section we find an expression for the Friedmann equation for the scale factor $a$ of the physical metric $g^{\text{eff}}_{\mu\nu}$, in terms of the functions $a$, $N$ and $\rho$. We recall that we will be setting $X=1$.

First, assuming $w\not=0$, we define the intermediate function $F(N)$ such that:
\begin{equation}\label{rhoprime}
\rho'\equiv F(N)N', 
\end{equation}
which can be obtained by taking the derivative of the constraint (\ref{Branch1constraint}) defining Branch I. The explicit expression for $F(N)$ is given by:
\begin{equation}\label{FofN}
F(N)= \left(\frac{2m^4}{ w\alpha\beta}\right)\frac{\left[(\alpha\beta_2-\beta\beta_1)+N(\alpha\beta_3-\beta\beta_2)\right]}{(\alpha+\beta N)^3}.
\end{equation}

On the other hand, by definition, we also have that:
\begin{equation}\label{Nprime}
N'=N(\mathcal{H}_f-\mathcal{H}_g).
\end{equation}
Combining this last equation with eq.~(\ref{rhoprime}) and the conservation eq.~(\ref{MatterConservation}) we get:
\begin{equation}\label{HHgHf}
\mathcal{H}_f=\mathcal{H}_g\left[1-\frac{3\mathcal{H}(1+w)\rho}{FN\mathcal{H}_g}\right],
\end{equation}
where we have also used that $p=w\rho$. By replacing this last equation into eq.~(\ref{Friedmann}) we find: 
\begin{equation}\label{EqHHg}
\mathcal{H}=\mathcal{H}_g\frac{(\alpha+\beta N)F}{(\alpha+\beta N)F+3\beta\rho(1+w)}.
\end{equation}
Next, we would like to eliminate the explicit dependence on $\mathcal{H}_g$ from the last equation, in favor of $a$, $N$ and $\rho$. In order to do this we define $\hat{H}_{i}$ for $i=(g,f)$ as: 
\begin{equation}
\hat{H}_{i}\equiv \sqrt{\mathcal{H}_{i}^2/(a_{i}X_{i})^2}>0, 
\end{equation}
and thus
\begin{equation}
\mathcal{H}_{i}=\pm \; \hat{H}_{i}a_{i}X_{i}, 
\end{equation} 
where there are in principle four combinations of signs to consider, but only two are relevant: if both $\mathcal{H}_{i}$ have the same sign, or different signs. Without loss of generality we consider only:
\begin{equation}\label{Hhat}
\mathcal{H}_{g}=\pm \; \hat{H}_{g}a_{g}X_{g}, \; \mathcal{H}_{f}= \; \hat{H}_{f}a_{f}X_{f}, 
\end{equation} 
and the other two cases will be different by an overall sign in $\mathcal{H}$ only. Explicitly, from the Friedmann equations for $g_{\mu\nu}$ and $f_{\mu\nu}$, given by eq.~(\ref{Friedg}) and eq.~(\ref{Friedf}), we get:
\begin{align}
\hat{H}_{g} & = \sqrt{\frac{1}{3}\left[\alpha \rho (\alpha+\beta N)^3+m^4(\beta_0+3N\beta_1+3N^2\beta_2+N^3\beta_3)\right]},\nonumber\\
 \hat{H}_{f} & = \sqrt{\frac{1}{3}\left[\beta \rho \frac{(\alpha+\beta N)^3}{N^3}+m^4(\beta_1 N^{-3}+3N^{-2}\beta_2+3 N^{-1}\beta_3+\beta_4)\right]}.\label{Hhats}
\end{align}
When replacing eq.~(\ref{Hhat}) into eq.~(\ref{EqHHg}), we get two possible expressions for $\mathcal{H}$, namely $\mathcal{H}_+$ and $\mathcal{H}_-$:
\begin{equation}\label{HXg}
\mathcal{H}_{\pm}=\pm X_{g}\frac{\hat{H}_{g}Fa}{(\alpha+\beta N)F+3\beta\rho(1+w)}.
\end{equation}

Finally, we express $X_g$ in terms of $a$, $N$ and $\rho$. From eq.~(\ref{XEffective}) we have that:
\begin{equation}\label{Xg1}
X_g=\frac{(\alpha+\beta N)-\beta N X_f}{\alpha}.
\end{equation}
but, eq.~(\ref{HHgHf}) can be rewritten as
\begin{equation}\label{EqXgXf}
\frac{\mathcal{H}_f}{\mathcal{H}_g}=\pm \frac{\hat{H}_f}{\hat{H}_g}N\left(\frac{X_f}{X_g}\right)=\left[1\mp \frac{3\mathcal{H}_\pm (1+w)\rho}{FN\hat{H}_ga(\alpha+\beta N)X_g}\right],
\end{equation}
where we have used eq.~(\ref{Hhat}) and eq.~(\ref{aEffective}). From here we can work out $X_f$ as a function of $X_g$ and the other background functions. When this result is replaced back into eq.~(\ref{Xg1})
we get for $X_g$:
\begin{equation}
X_g=\frac{(\alpha+\beta N)F\hat{H}_f N+3\mathcal{H}_\pm \beta\rho(1+w)(\alpha+\beta N)a^{-1} }{FN\left[ \alpha\hat{H}_f \pm \beta \hat{H}_g\right]}.
\end{equation}
Finally, combining this last equation with eq.~(\ref{HXg}), we find that the final Friedmann equation for $a$ is given by:
\begin{equation}\label{FinalFriedmann}
\mathcal{H}_{\pm}=\frac{ \hat{H}_{g} \hat{H}_{f} aFN(\alpha + \beta N)}{\pm \alpha N\hat{H}_{f}[(\alpha+\beta N)F+3\beta\rho(1+w)]+ \beta\hat{H}_g[N(\alpha+\beta N)F-3\alpha\rho(1+w)]},
\end{equation}
which corresponds to the Friedmann equation for the physical metric $g^{\text{eff}}_{\mu\nu}$ as a function of $\rho$, $a$ and $N$ (provided eq.~(\ref{FofN}) and eq.~(\ref{Hhats})).

From eq.~(\ref{FinalFriedmann}) we notice that the dependence of $\mathcal{H}_{\pm}$ on $\rho$ can be removed entirely by means of the constraint given by eq.~(\ref{Branch1constraint}), in which case we would get:
\begin{equation}
\mathcal{H}_{\pm}= a H_{\pm},
\end{equation}
where $H_{\pm}$ would be a function of $N$ only, and it would represent the Hubble rate in physical time. This last way of expressing $\mathcal{H}_{\pm}$ will be particularly useful throughout the paper, as a solution for $a$ can be found by analysing the behaviour of $N$ only.  

It is useful to show the explicit form of the brackets in the denominator as a function of $N$ only:
\begin{align}
(\alpha+\beta N)F+3\beta\rho(1+w) & = \frac{m^4}{\alpha\beta w (\alpha+\beta N)^2}\left[ \left(2\alpha\beta_2+\beta\beta_1(1+3w)\right)\right.\nonumber\\
& \left. +2N\left( \alpha\beta_2+ \beta\beta_2(2+3w)\right)+3\beta(1+w)\beta_3N^2\right],\\
N(\alpha+\beta N)F-3\alpha\rho(1+w) & =\frac{m^4}{\alpha\beta w (\alpha+\beta N)^2}\left[  -3\alpha\beta_1(1+w) - 2N\left( \beta\beta_1 + \alpha\beta_2 (2+3w)\right)\right. \nonumber\\
& \left. -N^2\left( 2\beta\beta_2 +\alpha\beta_3(1+3w)\right)\right],
\end{align}
where we have used the constraint given by eq.~(\ref{Branch1constraint}). From these last two expressions and eq.~(\ref{FinalFriedmann}) we can see that, if some of the parameters $\beta_n$, $\alpha$ or $\beta$ were negative and some positive, it would be likely to have a zero in the denominator for a finite value of $N$, which in general would generate a divergence in $\mathcal{H}$ at a finite time, making the solution for $a$ unviable. Even though there might be specific situations in which the parameters are chosen such that the divergence is avoided (with a compensating zero in the numerator), or the divergence occurs in the infinite future, we do not consider those highly specific cases further here. In addition, for negative values of the parameters there might also be violations of positivity of $\hat{H}_g$ and $\hat{H}_f$ at finite times, which would render those solutions unphysical as the Friedmann equations for the metrics $g_{\mu\nu}$ and $f_{\mu\nu}$ would become complex.
However, if instead all the parameters $\beta_n$, $\alpha$ and $\beta$ have the same sign, we can avoid a zero in the denominator, if $w>2/3$ and if we chose $\mathcal{H}_-$. This is the situation we consider in this paper. In particular, and without loss of generality, we assume all the parameters $\alpha$, $\beta$ and $\beta_n$ to be positive.

\section{Matter-dominated universe}\label{App:Dust}

In a matter-dominated universe the constraint given by eq.~(\ref{Branch1constraint0}) would become:
\begin{equation}\label{Bianchi1constraintMatter}
a_g^2 Z=0,
\end{equation}
where we have set $p=0$. This constraint can only be satisfied if $a_g=0$ ($N\rightarrow \infty$), or $Z=0$. However, since we are assuming all parameters to be positive, $Z=0$ can only be satisfied if $N=0$ when $\beta_1=0$. According to the previous results found during the radiation-dominated era, this constraint will be satisfied in the infinite past ($N=0$) and infinite future ($N\rightarrow \infty$) for the bouncing solution. Therefore, the matter-dominated solution is expected to describe the actual solution accurately only in these two limits. 
This last point can be easily seen for instance for late times. In this case the constraint will be satisfied by $a_g=0$, which means that $a=\beta a_f$, $X=X_f=1$, and thus $\mathcal{H}=\mathcal{H}_f$, where
\begin{equation}
\mathcal{H}^2=\frac{a^2}{3\beta^2}\left[\beta^4\rho_\text{m}+m^4\beta_4\right].
\end{equation}
Here we have used eq.~(\ref{Eqf}) and taken the $N\rightarrow \infty$ limit. At late times, when $\rho_\text{m}/m^4\ll 1$, the leading order term in this equation will coincide with that of eq.~(\ref{HlateDust}) and eq.~(\ref{HlateDust2}), but the next-to-leading order term will be different as effects of radiation become relevant. During early times, the constraint will be satisfied if $N=0$, and the same behaviour is found. Only the leading order term in the matter-dominated equation accurately describes the evolution of the universe. 

%--------------------------------------------------------------------------------------------------------------------------------------------
\bibliographystyle{JHEP}
\bibliography{RefMassiveBigravityDoubleCoupling}

\end{document}